\documentclass[conference]{IEEEtran}
\IEEEoverridecommandlockouts
\usepackage{cite}
\usepackage{amsmath,amssymb,amsfonts}
\usepackage{algorithmic}
\usepackage{graphicx}
\usepackage{textcomp}
\usepackage{cleveref}
\usepackage[utf8]{inputenc}
\usepackage{url}
\usepackage[singlelinecheck=false,justification=justified]{caption}
\usepackage{subcaption}
\usepackage{showexpl}
\usepackage{inconsolata}
\usepackage[export]{adjustbox}
\usepackage[table]{xcolor}
\usepackage{relsize}
\usepackage{float}

\newcommand{\smallertitle}[1]{\smaller{#1}}

\title{\smallertitle{Autonomous Electrochemistry Platform with Real-Time  Normality Testing  of  Voltammetry Measurements Using ML}
\thanks{This research is sponsored in part by the INTERSECT Initiative as part of the Laboratory Directed Research and Development Program and in part by RAMSES project of Advanced Scientific Computing Research program, U.S. Department of Energy, and in part by the Office of Basic Energy Sciences, Division of Materials Sciences and Engineering, U.S. Department of Energy, and is performed at Oak Ridge National Laboratory managed by UT-Battelle, LLC for U.S. Department of Energy under Contract No. DE-AC05-00OR22725.
The United States Government retains and the publisher, by accepting the article for publication, acknowledges that the United States Government retains a nonexclusive, paid-up, irrevocable, world-wide license to publish or reproduce the published form of this manuscript, or allow others to do so, for United States Government purposes. The Department of Energy will provide public access to these results of federally sponsored research in accordance with the DOE Public Access Plan (http://energy.gov/downloads/doe-public-access-plan).
}
}

\author{
\IEEEauthorblockN{
Anees Al-Najjar, Nageswara S. V. Rao}
\IEEEauthorblockA{{Computational Sciences and Engineering Division}\\
{Oak Ridge National Laboratory}\\ Oak Ridge, TN, USA\\
\{alnajjaram,raons\}@ornl.gov}
\and
\IEEEauthorblockN{
Craig A. Bridges, Sheng Dai}
\IEEEauthorblockA{
{Chemical Sciences Division}\\
{Oak Ridge National Laboratory}\\ Oak Ridge, TN, USA\\
\{bridgesca,dais\}@ornl.gov}
\and
\IEEEauthorblockN{
Alex Walters}
\IEEEauthorblockA{
{Energy Science Technology Division}\\
{Oak Ridge National Laboratory}\\ Oak Ridge, TN, USA\\
 waltersha@ornl.gov}
}

\begin{document}
\maketitle

\begin{abstract}
 Electrochemistry workflows utilize various instruments and computing systems to execute workflows consisting of electrocatalyst synthesis, testing and evaluation tasks. The heterogeneity of the software and hardware of these ecosystems makes it challenging to orchestrate a complete workflow from production to characterization by automating its tasks. We propose an autonomous electrochemistry computing platform for a multi-site ecosystem that provides the services for remote experiment steering, real-time measurement transfer, and AI/ML-driven analytics. We describe the integration of a mobile robot and synthesis workstation into the ecosystem by developing custom hub-networks and software modules to support remote operations over the ecosystem's wireless and wired networks. We describe a workflow task for generating I-V voltammetry measurements using a potentiostat, and a machine learning framework to ensure their normality by detecting abnormal conditions such as disconnected electrodes. 
We study a number of machine learning methods for the underlying detection problem, including smooth, non-smooth, structural and statistical methods, and their fusers. We present experimental results to illustrate the effectiveness of this platform, and also validate the proposed ML method by deriving its rigorous generalization equations.   
\end{abstract}

\section{Introduction}
\label{sec:introduction}
Instrument-computing ecosystems (ICE) that support Artificial Intelligence (AI)-automated experiments and computations are increasingly being deployed to improve scientific productivity in various science applications, including discovery of new materials, synthesis and characterization of new organic and inorganic compounds, and design and study of electrocatalysts and battery materials~\cite{al-najjar-xloop23, al2022enabling, enders2020cross, vescovi2022linking}. For chemistry applications of synthesis and characterization of compounds,
ICE examples include Materials Innovation Factory, University of Liverpool~\cite{mif}, the Autonomous Lab (A-Lab), Lawrence Berkeley National Laboratory (LBNL)~\cite{a-lab}, and the Matter Lab, University of Toronto~\cite{matter}. Several ICE workflows require significant computing and data resources~\cite{antypas2021enabling, tyler2022cross} to implement the services for remote controlling of physical instruments, real-time measurement streaming across facilities, and custom code execution on specialized computing platforms, all as part of AI-driven  orchestration for multiple rounds of experiments.

A testing workstation of electrochemistry ICE consisting of a flow reactor connected to a mass flow controller, syringe pump, peristaltic pump, fraction collector, and potentiostat has been previously developed for orchestrating remote experiments and analytics using real-time streaming of measurements~\cite{al-najjar-xloop23}. It also provides a machine learning (ML) method to check the normality of potentiostat measurements \cite{al-najjar-escience23} using an adhoc method designed for two classes.

In this paper, we propose a complete electrochemistry computing platform for a multi-site ICE that integrates synthesis instrumentation and delivery mobile robot, and supports real-time measurement transfer and sophisticated ML-driven analytics. It enables workflows spanning the synthesis of electrocatalysts, their real-time testing on multiple heterogeneous instruments, and ML computations on systems dispersed across multiple facilities.
We describe an ICE implementation by integrating a synthesis workstation that automates compound preparation of a catalyst and a mobile robot that delivers the catalyst in a vial to the fraction collector, which requires two different networking and software eSolutions.
We design separate custom hub-networks to connect these instruments to a  gateway computer, and develop software modules to integrate them into ICE. 
The Kuka mobile robot is integrated over a dedicated wireless network using a media converter. The Chemspeed SwingXL synthesis workstation is integrated over a wired Ethernet hub-network using its custom software.

For a successful execution of an automated electrochemistry workflow lasting for days  to weeks, it is critical to  ensure that accurate I-V voltammetric measurements are collected by the potentiostat connected to the flow reactor via electrodes. By utilizing a ``normal'' shape of I-V voltammogram profile indicated by Cyclic Voltammetry (CV) electrochemistry technique, we develop a custom ML method to  detect normal and abnormal conditions such as electrode or instrument disconnection and related failures\footnote{A demonstration of a limited version of such capability is presented as a poster \cite{al-najjar-escience23} using an adhoc ML method, and this paper presents a systematic study of a variety of ML methods (since no single best method exists) to develop a fused-classifiers method with rigorous performance guarantees.}.
This ML method utilizes I-V measurements collected under normal and other conditions for training composed of two steps: (i) {\it feature estimation:}
the Gaussian process regression (GPR) method is used to extract a 10-d feature from I-V measurements, and 
(ii) {\it classification:} the extracted features are used by a classifier to determine the normality.
The underlying normality checking is a complex task since I-V measurements sets have a varying number points in hundreds, which makes a direct application of known ML methods impractical, and our method addresses it by reducing them to fixed size 10-d feature vectors. 
Furthermore, there is no single best classification method that uses finite samples \cite{DGL96} as a consequence of  mostly unknown measurement and sensing error distributions of the potentiostat. 
In response, we systematically study a variety of ML methods,  including smooth, non-smooth, structural and statistical, and their fusers that combine their strengths. Then, we identify a fused-classifiers method based on experimental and analytical performance analysis. 
Specifically, for both regression-based feature estimation and subsequent classification, we derive the generalization equations that establish their effectiveness beyond the training performance \cite{Vapnik98}.

\begin{figure*}[t]
\centering
\includegraphics[width=\textwidth,height=2.5in]{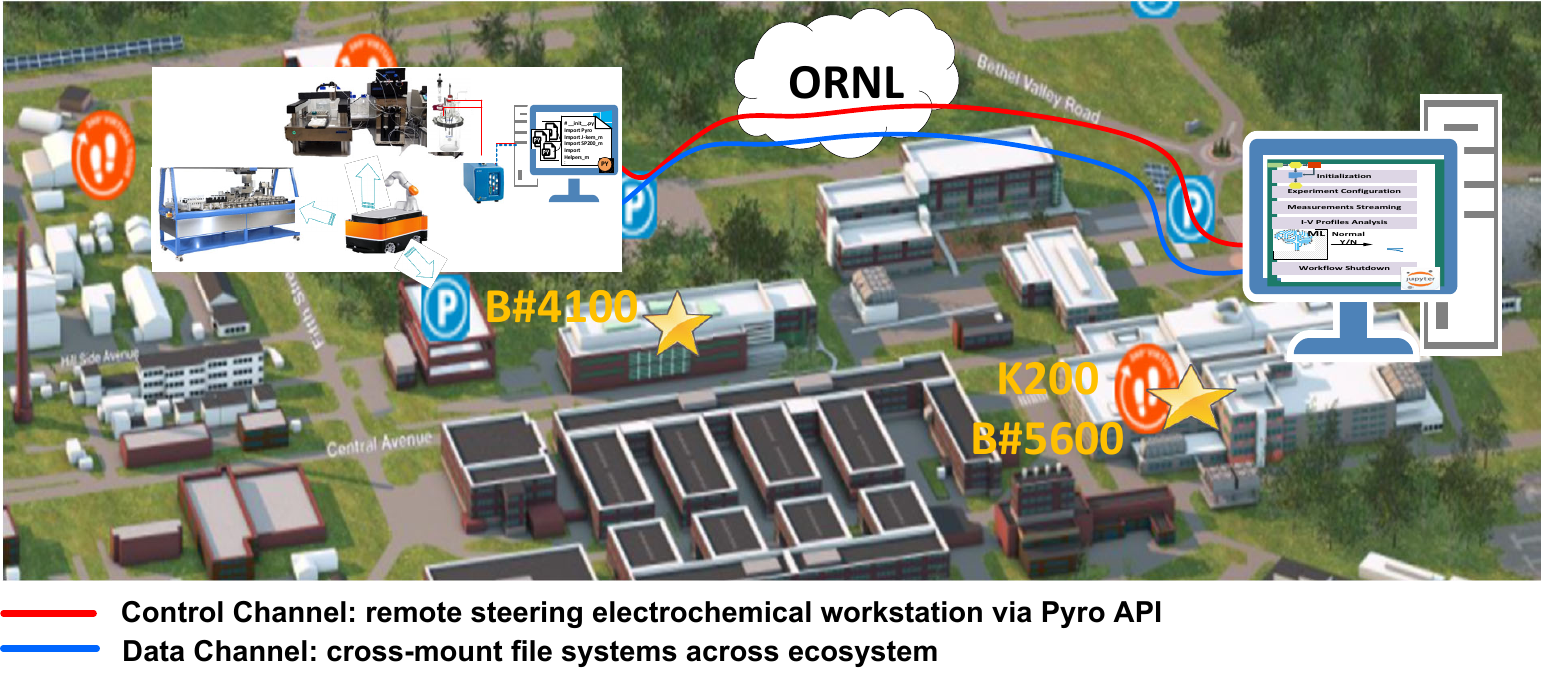}
\caption{{ORNL Electrochemistry ecosystem consists of instruments and computing facilities connected over networks.}}
\label{fig:electrochemistry_Ecosystem}

\vspace*{-0.2in}
\end{figure*} 

We present experimental results illustrating the effectiveness of the ecosystem integration tasks and the normality detection method.
The synthesis operation is activated from the gateway computer to demonstrate the task of preparing a vial of electrochemical compound for the robot pickup.
The Kuka robot is activated from the gateway computer to deliver the vial from the synthesis workstation to the fraction collector located across the laboratory room.
The proposed ML method is assessed using measurements collected under four different conditions, and is shown to correctly determine normal and abnormal conditions.

The organization of this paper is as follows.
Related works are  briefly presented in Section \ref{sec:Related_work}.
The autonomous chemistry laboratory and electrochemistry workflows are briefly described in Section \ref{sec:Electrochemistry_Ecosystem_n_Workflows}.
The ML framework is presented in Section \ref{sec:ML_Framework}.
The study of multiple classifiers and their fusers is presented in Section \ref{sec:ML_Classifier_Design_n_Analysis}. 
Conclusions and future research directions are described in Section  \ref{sec:conclusions}.

\section{Related work}
\label{sec:Related_work}

The A-Lab~\cite{a-lab} is reported to develop inorganic solid-state materials by automating various hardware and software platforms of robots, incorporating chemistry instruments and software tools utilized for different analyses. The Matter Lab~\cite{matter} at The University of Toronto  is also developing self-driving chemistry experiments using multiple pieces of equipment for synthesis and characterization. In addition, MIF~\cite{mif} at the University of Liverpool  is also another self-driving laboratory to support automated chemistry and material experiments. Besides these main contributions, other related works explain experiment automation via customizable workflows with Labview~\cite{dave2022autonomous,gerroll2023legion}, local experiment control~\cite{joress2022development,laws2024autonomous}, experiment orchestration using specific software control --as in case with ChemOS~\cite{seifrid2022autonomous}, or distributed framework~\cite{rahmanian2022enabling}. However, these solutions generally have limited software and network design capabilities to support real-time and autonomous orchestration of chemistry experiments across multi-site ecosystems. Another related contribution is the hardware platform in \cite{rial2024automated}, which automates electrochemical flow reactions with specialized electrochemistry instruments using a directly connected control system. This platform is mainly designed to steer the electrochemistry experiments locally with limited workflow orchestration and computing capabilities, and it does not support the remote orchestration of the electrochemistry setup and the analytics on high-performance remote computing systems as our proposed platform does. Our platform considers autonomous electrochemistry workflow from synthesis production (conducted on SwingXL) to reaction testing (performed on the electrochemistry testing workstation), where the catalyst is transferred between these workstations via Kuka robotic platform.

Emerland Cloud Lab~\cite{emeraldcloudlab} is a mature cloud-based commercial project for science research and development in multiple areas, including chemistry, biotech, pharmaceutical, and materials research. It includes a variety of science instruments autonomously orchestrated by the vendor software API that also orchestrate the compute services deployed over the cloud. Although there is no clear perception of how the lab ecosystem is designed, our proposed electrochemistry ecosystem encompasses multiple facilities of instruments and HPC systems located at different facility domains, which requires a notional software and hardware design for integrating instrument and computing systems and support autonomous electrochemistry workflows across multiple facilities.

Current workflow frameworks (such as Pegasus, Merlin, Fireworks, and ExaWorks) and interconnected compute services platforms (such as Globus Compute~\cite{chard2020funcx,bauer2024globus}), are mainly concerned with managing interconnected computational tasks over a distributed computing environment, wherein the data involved in these tasks are presumed to be \textit{in-situ} offloaded or \textit{generally} integrated as part of the workflows~\cite{pegasus,merlin,fireWorks,exaWorks}. They prescribe the mapping of compute and data services across computing nodes and the potential decisions made after the execution of certain tasks. However, these workflow frameworks have not fully addressed the real-time and autonomous orchestration requirements of science experiment steering with instrument control and computations over ICE. 

Such combined capabilities have been addressed in the proposed contribution in this paper, as well as in our previous work that enables autonomous orchestration of instrument control and data transfer and real-time analytics at HPC systems~\cite{al2022enabling,al-najjar-xloop23}. This is achieved through addressing software and network infrastructure design requirements across instrument-computing ecosystems with multiple facilities~\cite{smartcomp23-cyber-framework} to steer experiments and support real-time AI-driven analytics remotely, example illustrated in~\cite{al-najjar-escience23}. However, the proposed platform integrate new capabilities into the electrochemistry ICE -- for synthesizing the catalyst and transferring it across the facility-- and more fine-tuned ML for CV experiments carried out across multi-facility ecosystems, which has not been presented previously.


\section{Electrochemistry  ICE and Workflows}
\label{sec:Electrochemistry_Ecosystem_n_Workflows}

\begin{figure}[t]
\centering
\includegraphics[width=0.4\textwidth,height=2.5in]{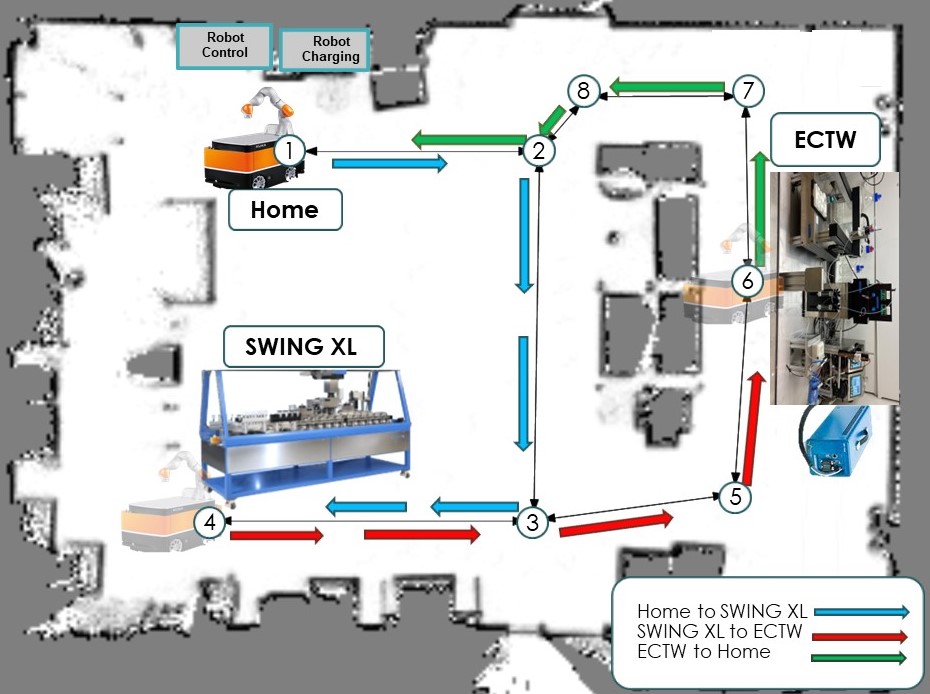}
\caption {{ACL facility map generated by Kuka robot using its LiDAR range sensor system, which is used for navigation.}} 
\label{fig:acl_map_for_kuka}
\vspace{-0.5cm}
\end{figure}

\begin{figure*}[t]
\centering
\includegraphics[width=0.6\textwidth,height=2.8in]{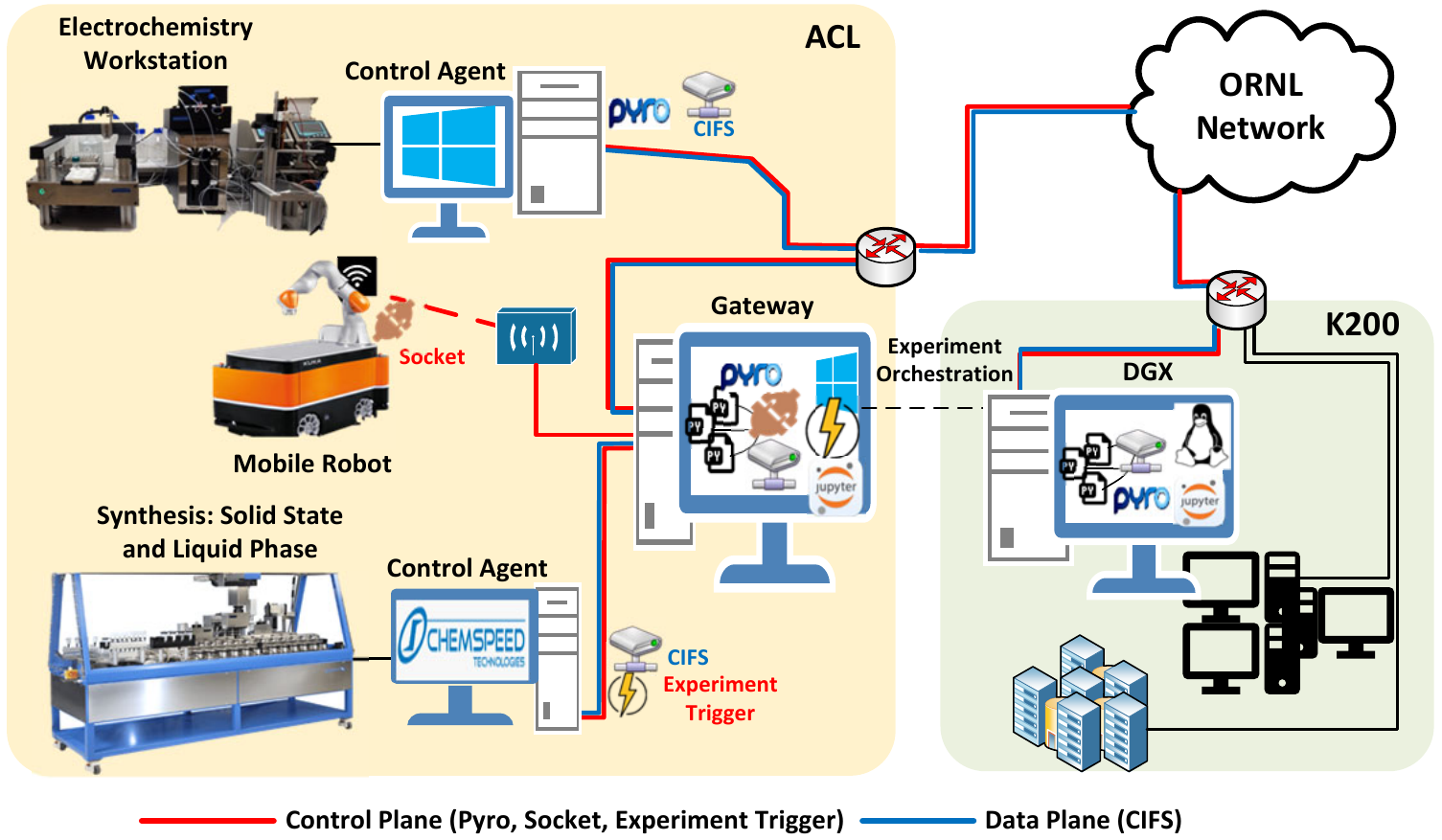}
\caption {Electrochemistry ecosystem integration design}
\label{fig:Electrochemistry_Ecosystem_Integration_Design}
\vspace*{-0.15in}
\end{figure*}

\begin{figure}[t]
\centering
\includegraphics[width=0.4\textwidth,height=2.2in]{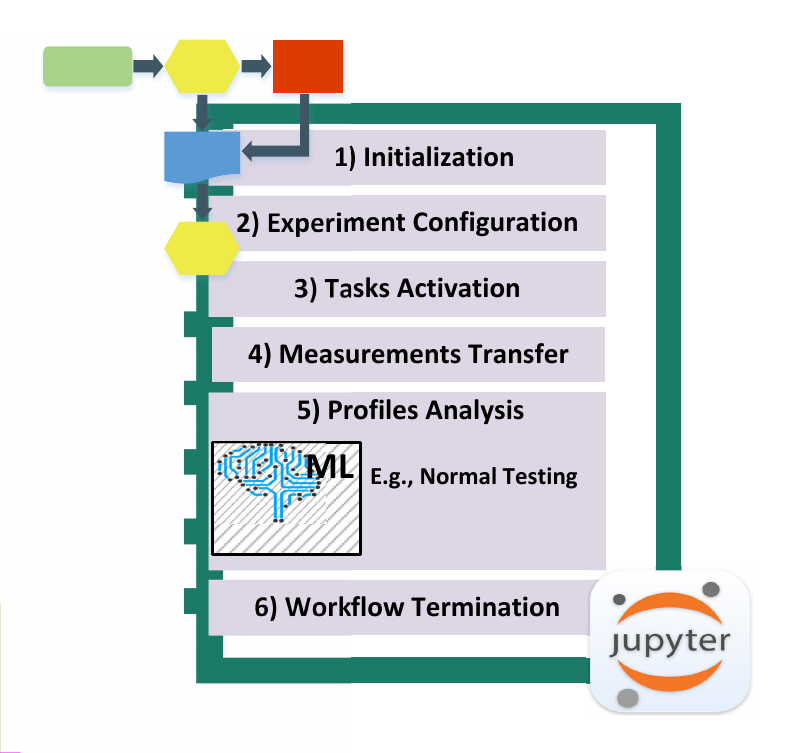}
\caption{{Electrochemistry workflow components}}
\label{fig:electrochemistry_workflow}
\end{figure}

The autonomous electrochemistry experiments require  real-time  remote orchestration of workflows across the Oak Ridge National Laboratory (ORNL) ecosystem shown in Fig.~\ref{fig:electrochemistry_Ecosystem}. The electrochemical testing workstation portion of the ecosystem with a limited normality detection capability was described in~\cite{al-najjar-xloop23,al-najjar-escience23}. 
In this paper, we describe the complete ICE augmented with a synthesis workstation, Kuka mobile robot, and a comprehensive ML pipeline for real-time normality detection under different testing experiment failures modes.

\begin{figure*}[t]
\centering
\begin{subfigure}[c]{0.23\textwidth}
   \includegraphics[width=\textwidth,height=1.5in]{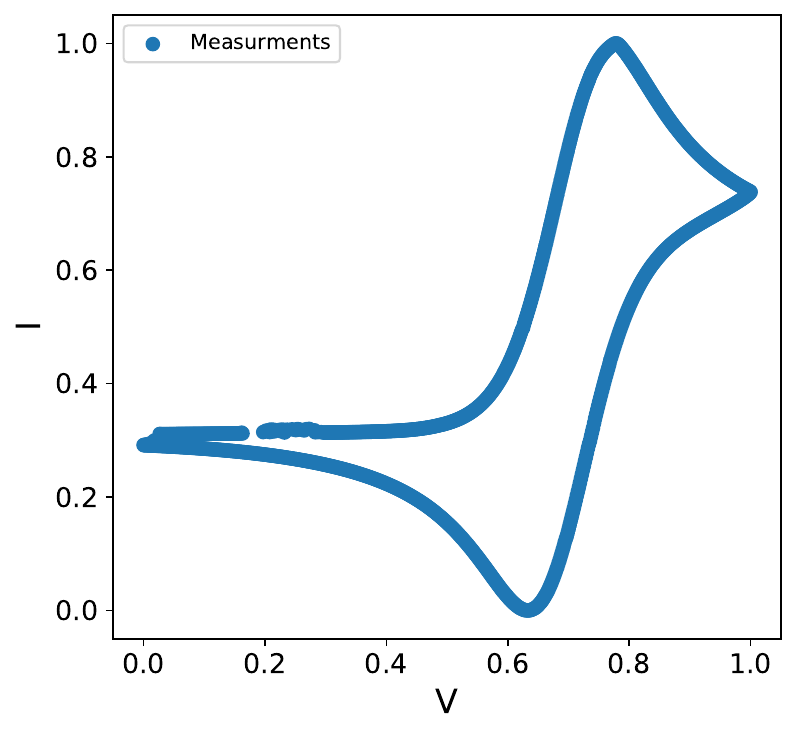}
\caption{{normal I-V measurements}}
\label{subfig:connected_normal}
\end{subfigure}
\hfill
\begin{subfigure}[c]{0.23\textwidth}
        \includegraphics[width=\textwidth,height=1.5in]{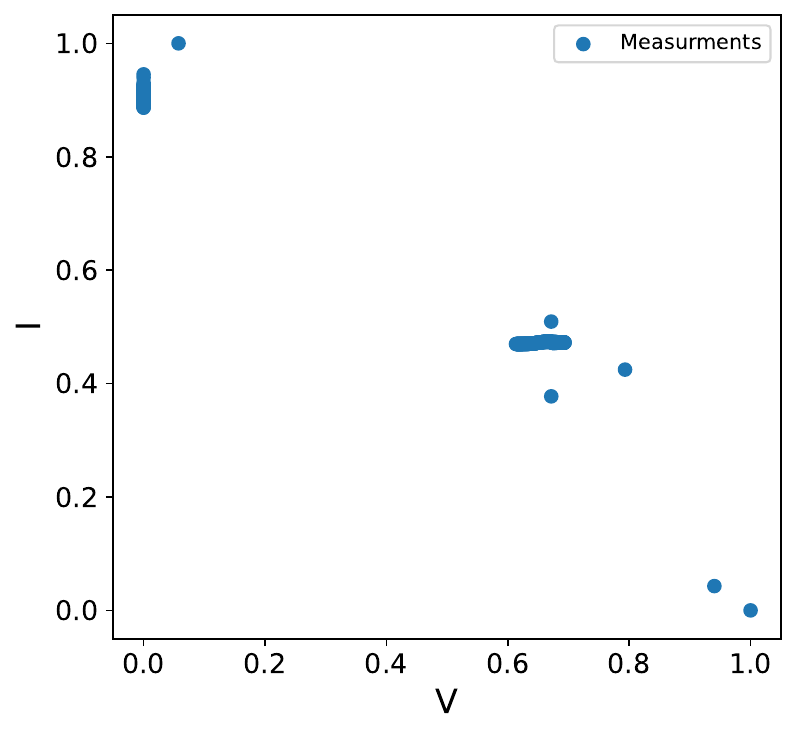}
\caption {{reference disconnected}}
\label{subfig:disconnect_ref}
\end{subfigure}
\hfill
\begin{subfigure}[c]{0.23\textwidth}
   \includegraphics[width=\textwidth,height=1.5in]{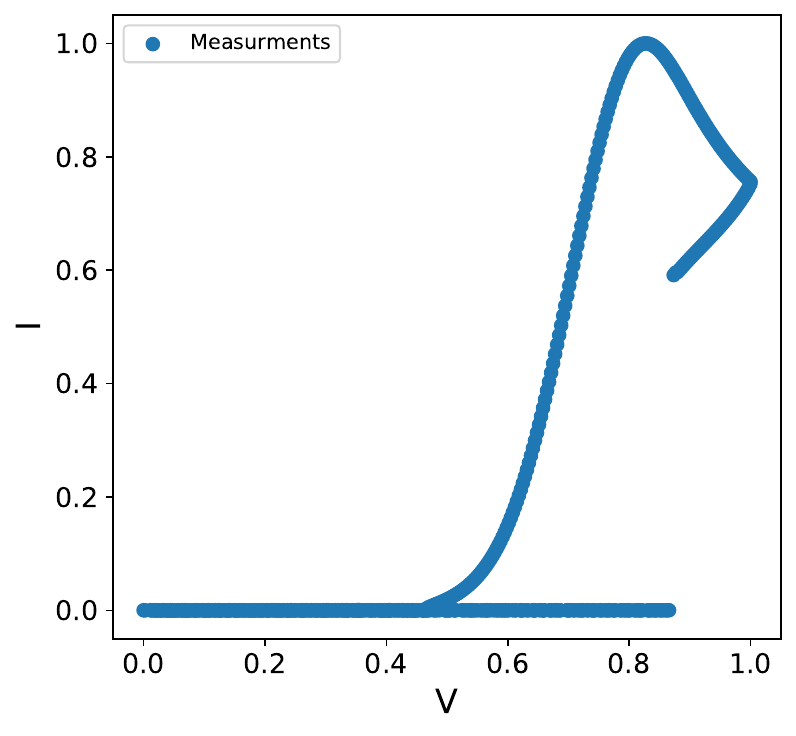}
\caption{{working disconnected}}
\label{subfig:disconnect_working}
\end{subfigure}
\hspace*{2em}
\begin{subfigure}[c]{0.23\textwidth}
        \includegraphics[width=\textwidth,height=1.5in]{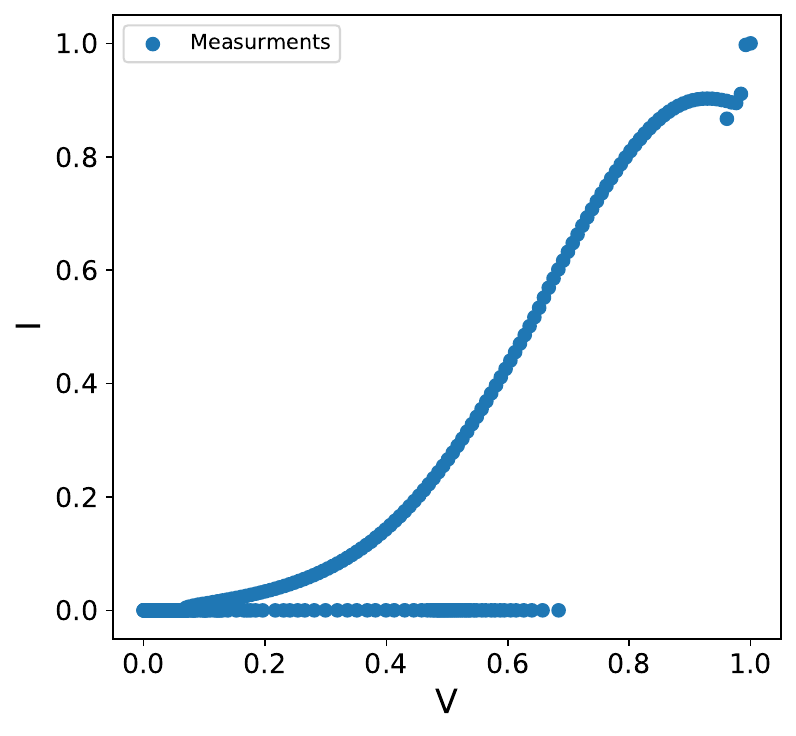}
\caption{{counter disconnected}}
\label{subfig:disconnect_counter}
\end{subfigure}
\hfill
\caption{{I-V measurements under normal, and reference, working, counter disconnected experimental conditions.}}
\label{fig:I-V measurements}
\vspace*{-0.15in}
\end{figure*}

\subsection{Cross-Facility Electrochemistry Ecosystem}
\label{subsec:Cross-facility_Electrochemistry_ecosystem}

The ORNL ICE spans computing and instrument platforms deployed at two facilities that are connected to distinct network domains, as shown in Fig.~\ref{fig:electrochemistry_Ecosystem}.

\subsubsection{Autonomous Chemistry Laboratory}
\label{subsubsec:Autonomous Chemistry Laboratory}
The Autonomous Chemistry Laboratory (ACL) at ORNL supports R\&D in the synthesis and testing of organic and inorganic compounds in liquid or solid state. It is utilized to investigate testing of catalyst compounds by utilizing multiple specialized instruments and robotic workstations.
The SwingXL Robotic synthesis workstation (from Chemspeed) is a recent addition that enables the preparation of liquid catalyst solutions with prescribed automated composition. The prepared vial is positioned to be picked up by  Kuka mobile robot, 
which transfers it to the testing workstation along the pre-trained path depicted in Fig.~\ref{fig:acl_map_for_kuka}. 
The figure illustrates a 2D scan of ACL taken by the Kuka on-board LiDAR (Light Detection and Ranging) sensors, which is used as a map of the workspace. The ACL workplace includes numbered locations, called nodes, which the Kuka mobile platform navigates to while following the arrows, called edges. The blue route shows the path the mobile platform takes from its home location (1) to the SwingXL pick up location (4). The red route shows the path from the SwingXL to the electrochemistry testing workstation (ECTW) (6), and the green route shows the path back to home (1). 

The robot training and calibration on the selected path is programmed, and its actions are triggered from a remote computing system. These steps include moving from the home position towards SwingXL to pick the vial with synthesized liquid compound, and place it at the prescribed location on the faction collector of at the electrochemistry testing workstation. 
The testing workstation consists of specialized chemistry instruments of a syringe pump, fraction collector, and mass-flow controller, which together deliver a required amount of liquid from the vial to the flow reactor. Then, a potentiostat connected to the flow reactor via electrodes collects the voltagram I-V measurements (described in detail in the next section).

\subsubsection{K200 Facility}
\label{subsubsec:K200_Data_and_Compute_Facility}

This facility hosts high-performance computing workstations and clusters, as well as storage and server systems. For instance, it hosts  Nvidia DGX workstations with multiple GPU and CPU units, such as DGX-1 with eight GPUs and twenty CPUs which is operated by Linux Ubuntu 20.04LTS. The workflow proposed in this paper is executed from DGX-1 over ORNL ecosystem to autonomously orchestrate the workflow control and compute services.

\subsection{Integration of SwingXL and Kuka Robot} 
\label{{subsec:Ecosystem_Integration}}

The ACL platforms, SwingXL robotic synthesis workstation, Kuka mobile robot, and  electrochemistry testing workstation, are locally controlled using specialized APIs embedded in their control agents. 
They require integration solutions to enable autonomous workflow control and real-time measurement transfer and analytics on remote computing systems.  We integrate the electrochemistry platforms at ACL using  custom designed hub-networks connected to the back of ACL gateway computer, which in turn is connected to ORNL network, as shown in Fig.~\ref{fig:Electrochemistry_Ecosystem_Integration_Design}. 
The network channels are utilized to transfer data and control messages over respective planes managed by services and control agents.

The SwingXL system is integrated via a custom local network connected to the back of ACL gateway computer. We mounted the experiment file system at the control agent using CIFS protocol on the gateway computer to be accessed over the local network. The synthesis experiment at SwingXL is triggered from the gateway computer by moving the experiment configuration and metadata files into the experiment directory at the control agent, which are predefined in the workflow created by the Chemspeed control software. 

Kuka robot, on the other hand, requires different software and hardware integration design. It requires WiFi connection which mandates ecosystem access over wireless infrastructure. ORNL organization policy does not allow dual-homed wired and wireless networks on Windows workstations, and thus ACL gateway computer can not be connected to Kuka robot and ORNL network at the same time. We overcome this limitation by plugging a WiFi media converter extension to Ethernet back end of the gateway computer. Thus, Kuka wireless connection is converted into a wired connection at the gateway. 
For software integration, Kuka uses Java API embedded in the platform to program its functionalities, such as platform movements inside ACL for catalyst transfer. Our proposed workflow, however, does not support the Java front end interface since it orchestrates the workflow tasks across multiple instrument and compute systems using Python-based Jupyter Notebook. Hence, we developed a cross-platform solution to control Kuka steps programmed in Java from the gateway system using socket network programming libraries for implementing server-client modules. The back end socket module is implemented using Java and deployed on Kuka robot that runs as a server and wraps multiple methods associated with Kuka movements around ACL as objects to be accessed across the ecosystem network. The front end socket modules are implemented in Python and deployed on the remote computing system for calling the Java server objects.

Integration of the electrochemistry testing workstation to test catalyst solution in the vial provided by SwingXL, is described in our previous work~\cite{al-najjar-xloop23}. The testing workstation consists of specialized chemistry instruments of syringe pump, fraction collector, and mass-flow controller, all connected to a vendor control system (single-board computer) from J-Kem, which is accessed to control the setup via remote API developed by ORNL\footnote{https://github.com/aneesalnajjar/electrochemistry}. Briefly, the control commands of instruments of J-Kem setup and Bio-Logic potentiostat are wrapped as Pyro objects~\cite{pyro} at the control agent and accessed by their peers of Pyro clients from DGX system at K200. The Jupyter Notebook at DGX autonomously orchestrates the remote experiment at the workstation, and transfers and analyzes reaction profiles in real time. The profiles are made available across the ecosystem via data plane service using Common Internet File System (CIFS) protocol for cross-mounting file systems.


\subsection{Electrochemistry Workflow}
\label{subsec:Electrochemistry_Workflow}

An electrochemistry workflow may be orchestrated from a remote computing system located at different facility across the ecosystem, for instance, DGX at K200 shown in Fig.~\ref{fig:electrochemistry_Ecosystem}. It is executed in real-time in steps:
synthesizing the catalyst at the SwingXL, 
transferring it by Kuka mobile robot, testing the catalyst at the electrochemistry workstation, and collecting and transferring results for analysis at the remote computing system, as explained in Fig.~\ref{fig:electrochemistry_workflow}. Prior to starting the workflow, \textit{human-in-the-loop} is required to examine the operational status of ACL platforms and their instruments, as well as ensuring the chemical compounds are available in the SwingXL as needed to prepare the catalyst.

The workflow is initialized over the control plane via server-client communication between the control agents at ACL and remote computing systems by activating server modules (\textbf{Step 1}). Different communication protocols are utilized based on the platform setup and instrument APIs. Pyro library is utilized to communicate with the control agent of the electrochemistry testing workstation, while socket library is utilized to communicate with Kuka robot across the ecosystem network.

ACL workstations are configured with experiment metadata to be interactively maintained as part of the workflow orchestrated from the remote systems (\textbf{Step 2}). The metadata messages are conveyed using Pyro modules to configure the electrochemical workstation, and socket modules to configure Kuka moves. Configuring SwingXL is done by using a text file that includes synthesis steps and the instruments' actions to control the catalyst process. The configuration file is sent from the remote computer into a directory specified by the ChemSpeed control software on the control agent. Meanwhile, configuring the electrochemistry testing workstation includes specifying the potentiostat firmware and the electrochemistry technique binary files, metadata for configuring the firmware and experiment technique, serial connection parameters between control agents and the instruments, as well as data plane configurations related to file system mounts.

Thereafter, the workflow tasks are activated and orchestrated from remote computing systems (\textbf{Step 3}). The step comprises executing synthesis tasks at the SwingXL, including making the catalyst, making a slurry with the catalyst, and depositing and drying it. When the catalyst is ready, it is placed in a holder at the SwingXL pickup location, whereupon the Kuka robot picks it up and moves it across the lab --following the steps shown in Fig.~\ref{fig:acl_map_for_kuka}-- to the electrochemistry testing workstation. Then, the sample is placed at the fraction collector, and the syringe pump spouts the liquid sample into the flow reactor. 
Once pumping the liquid to the cell is completed, the redox testing process at the cell is triggered, and the measurements are made available at the remote system across the ecosystem through for real-time analytics (\textbf{Step 4}). After the oxidation/reduction cycling process is completed and the profiles are successfully collected, science modules can access the measurements and perform various domain-related computations (\textbf{Step 5}). An example of such applications is validating these profiles' normality before advanced domain computations. 
Finally, when all the scientific campaign tasks are completed, the workflow is shut down by disconnecting the Pyro and socket communication between the control agents and remote computing systems that orchestrate the workflow across ORNL ecosystem (\textbf{Step 6}).

While testing the synthesized catalyst at the electrochemistry testing workstation is orchestrated from the DGX system at K200 (explained in~\cite{al-najjar-xloop23}), other workflow tasks associated with the synthesis and catalyst transfer using SwingXL and Kuka robot platforms, respectively, are autonomously orchestrated by a Jupyter notebook from the gateway system at ACL.

\section{ML Framework for Volatammetry}
\label{sec:ML_Framework}
\begin{figure}[t]
\centering
\includegraphics[width=.4\textwidth,height=2in]{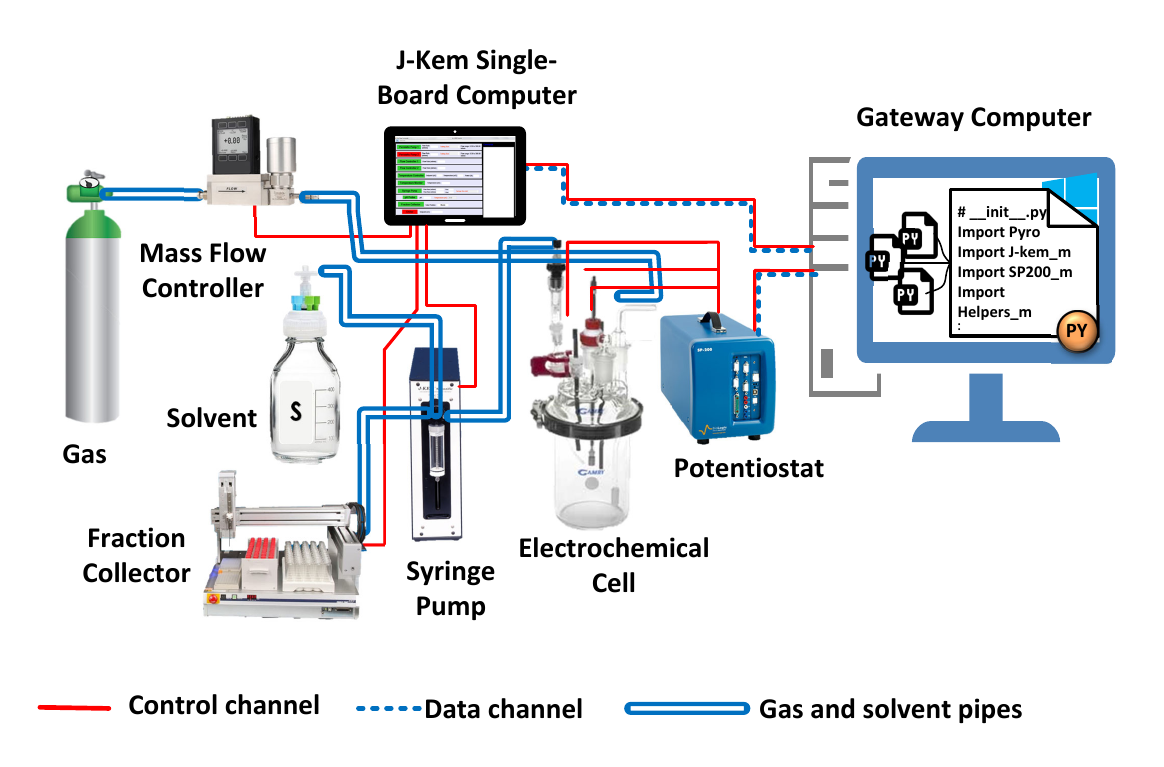}
\caption {{Electrochemistry testing workstation \cite{al-najjar-xloop23}}.}
\label{fig:electrochemistry_workstation}
\vspace*{-0.2in}
\end{figure}

The electrochemistry workflows are expected to be autonomously executed for days to weeks, wherein I-V measurements are 
repeatedly collected at the testing workstation an shown in Fig.~\ref{fig:electrochemistry_workstation}.
Specifically, these I-V measurements are collected by the potentiostat by sweeping the applied voltage up and down while measuring the resultant current using CV experiments.
The data are then transferred and analyzed remotely to identify the next composition of the catalyst to be prepared in SwingXL, and thus drive the critical workflow computations consisting of domain and AI codes.
We now describe a framework for ensuring the normality of these I-V measurements by utilizing ML method that detects normal conditions and abnormal conditions such as disconnected electrodes.

\begin{figure*}[h]
\centering
\begin{subfigure}[c]{0.23\textwidth}
   \includegraphics[width=\textwidth,height=1.3in]{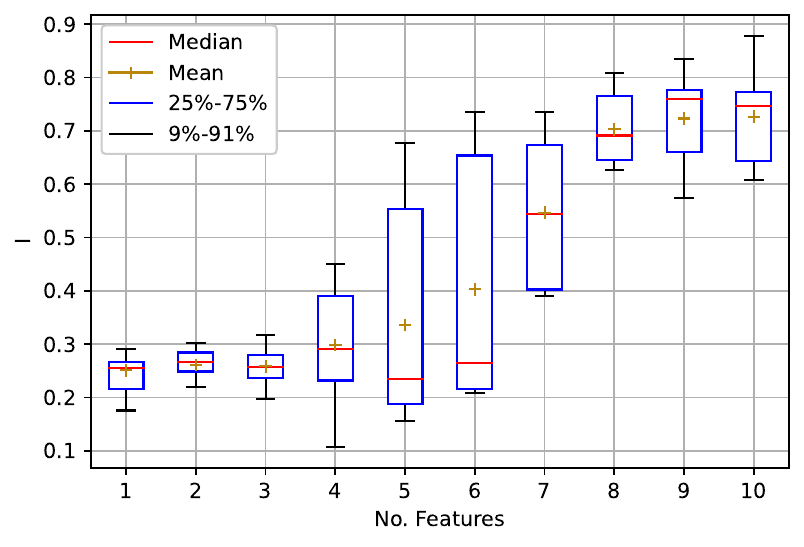}
\caption{{normal condition}}
\label{subfig:testing_normal_iv_profile}
\end{subfigure}
\hfill
\begin{subfigure}[c]{0.23\textwidth}
        \includegraphics[width=\textwidth,height=1.3in]{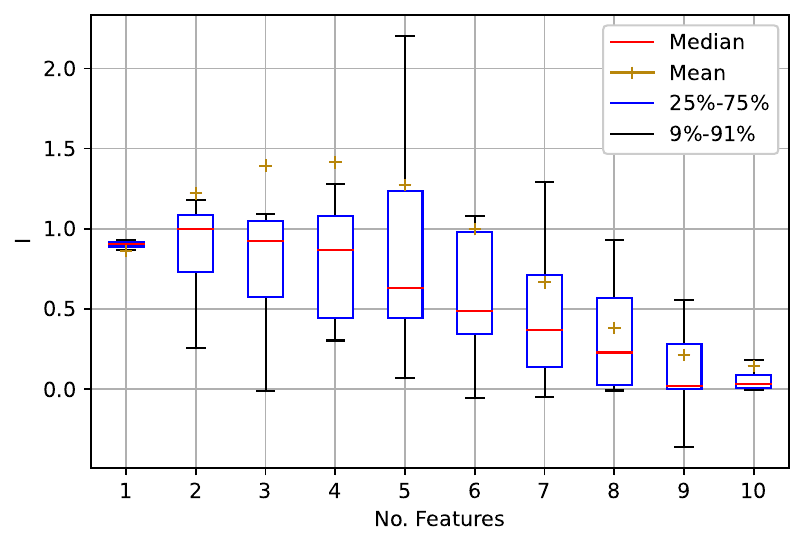}
\caption {{dis. reference electrode}}
\label{subfig:classifying_testing_normal_iv_profile}
\end{subfigure}
\hfill
\begin{subfigure}[c]{0.23\textwidth}
   \includegraphics[width=\textwidth,height=1.3in]{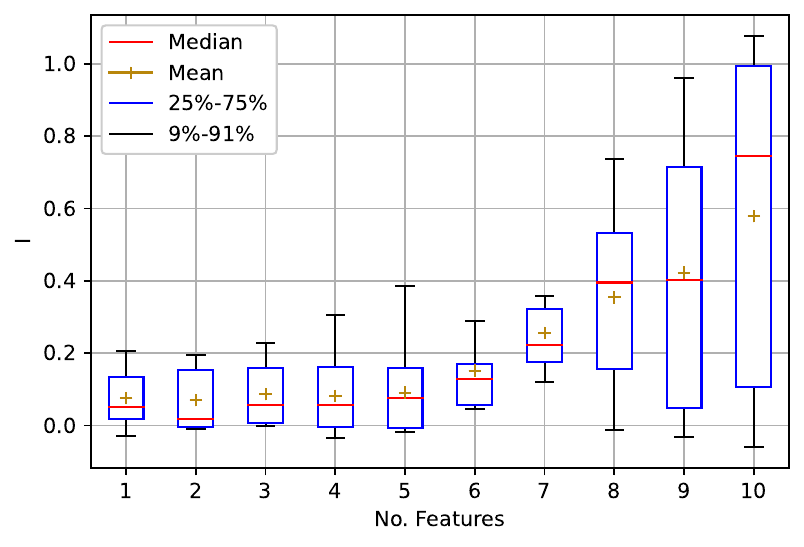}
\caption{{dis. working electrode}}
\label{subfig:testing_invalid_iv_profile}
\end{subfigure}
\hspace*{2em}
\begin{subfigure}[c]{0.23\textwidth}
        \includegraphics[width=\textwidth,height=1.3in]{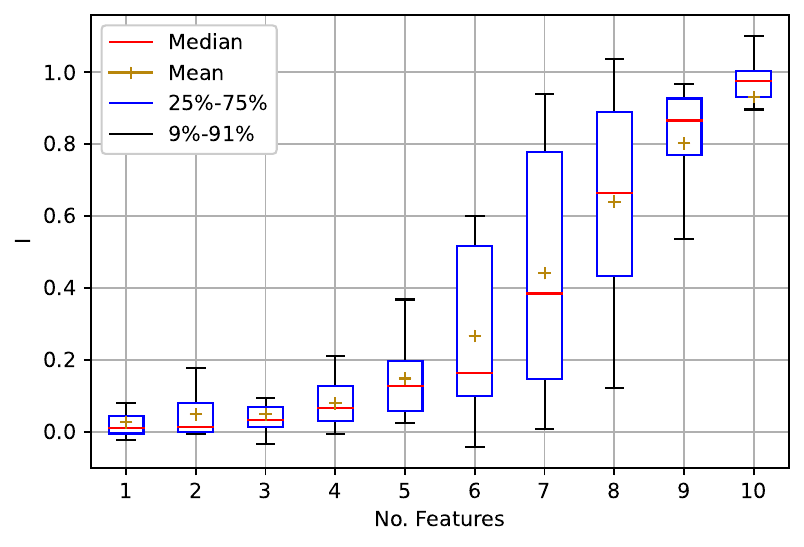}
\caption{{dis. counter electrode}}
\label{subfig:classifying_testing_invalid_iv_profile}
\end{subfigure}
\hfill
\caption{{GPR features of CV datasets under normal condition, and disconnected (dis.) reference, working and counter electrodes.}}
\label{fig:ml_results}
\vspace*{-0.15in}
\end{figure*}

\subsection{Electrochemical Technique: Cyclic Voltammetry} 
\label{subsec:Cyclic_Voltammetry}

The CV method used by the potentiostat results in a varying number of individual data points during the I-V measurements (typically in hundreds) but with a specific geometric shape; while the shape can vary to some extent depending upon the nature of the redox active ions being investigated, significant deviations can occur that reflect abnormal conditions.
Fig.~\ref{fig:I-V measurements} shows I-V profiles of a standard solution of the metal complex ferrocene in acetonitrile, with the supporting electrolyte salt tetrabutylammonium triflate at 0.1M concentration (more details in~\cite{al-najjar-xloop23}). The data are collected using a platinum working and counter electrode, and a silver reference electrode. It shows the normal I-V voltammogram profile (Fig.~\ref{subfig:connected_normal}), and abnormal profiles due to disconnection of the electrodes, namely, reference (Fig.~\ref{subfig:disconnect_ref}), working (Fig.~\ref{subfig:disconnect_working}), or counter (Fig.~\ref{subfig:disconnect_counter}). 
The profiles are analyzed in real-time at the remote system once the reaction at the electrochemistry workstation is completed.


\begin{figure}[h]
\begin{center}
\includegraphics[width=0.5\textwidth]{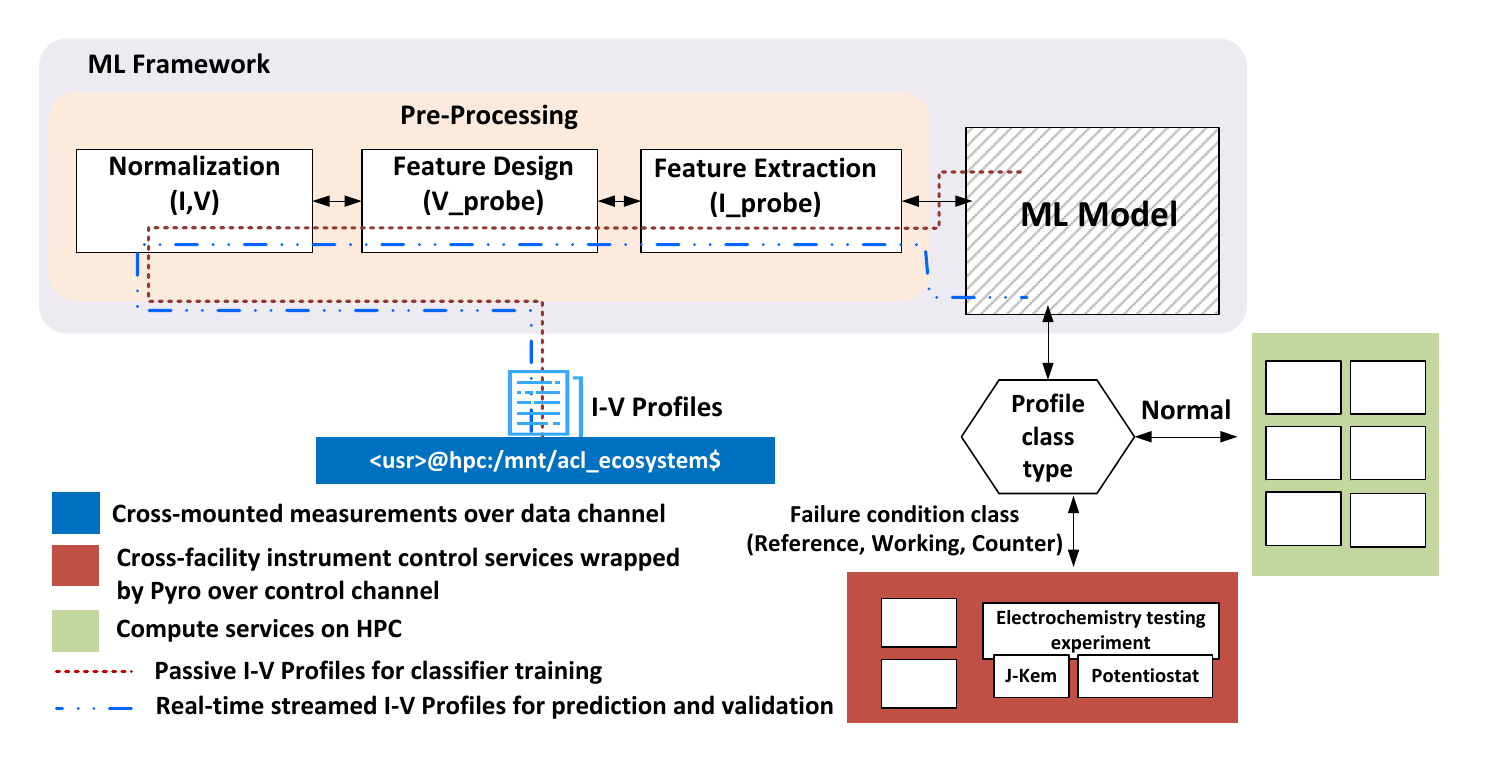}    
\end{center}
\caption{ML framework for I-V normality testing integrated into electrochemistry workflow.} 
\label{fig:IV_Profiles_ML_Framework}
\vspace*{-0.1in}
\end{figure}

\subsection{Cyclic Voltammetry Dataset}
\label{subsec:dataset}
The dataset is obtained  by the CV  technique using a solution of the metal complex ferrocene in acetonitrile, with the supporting electrolyte salt tetrabutylammonium triflate at 0.1M concentration. 
The I-V measurements --- including Potential (V) and Current (I), as well as Timestamp (ms)--- are collected in multiple real-time experiments conducted by the workflow explained in~\cite{al-najjar-xloop23}. We collected 60 (I-V) profiles under normal conditions and failures corresponding to disconnected  reference, working, and counter electrodes; they represent four distinct CV experiment classes, as detailed in Table~\ref{tab:CV_Dataset_Classes}. 
We utilize these measurements to train different ML models and assess their accuracy as described in next section.

\begin{table}[htbp]
  \centering
  \caption{CV Dataset Classes}
    \begin{tabular}{|c|c|c|}
    \hline
    \rowcolor[rgb]{ .851,  .851,  .851} \textbf{Class ID} & \textbf{Class Name} & \textbf{Count} \\
    \hline
    \cellcolor[rgb]{ .851,  .851,  .851} 1& normal      & 14 \\
    \hline
    \cellcolor[rgb]{ .851,  .851,  .851} 2& reference     & 16 \\
    \hline
    \cellcolor[rgb]{ .851,  .851,  .851} 3& working      & 16 \\
    \hline
    \cellcolor[rgb]{ .851,  .851,  .851} 4& counter      & 14 \\
    \hline
    \end{tabular}%
  \label{tab:CV_Dataset_Classes}%
\end{table}%

\vspace{-0.1in}
\subsection{Feature Analysis}
\label{Feature_Analysis}

The I-V measurements typically consist of a varying number of points (typically in 100s)  but with a clearly defined shape, and we extract their 10-d signatures to be used as input to the classifier. This 10-d feature vector is designed to capture
the I-V profile shape by fitting  a continuous regression curve with  V and I (both normalized to range (0,1)) as independent and dependent variables, respectively, and extracting 10 regression points at chosen V-values. It is a 10-d vector with I-values computed at fixed probe points in V-space by fitting a GPR model to measurements. The GPR fits to I-V measurements and corresponding feature vectors at probe voltages under normal and abnormal conditions are shown in Fig.~\ref{fig:ml_results}(a)-(d) for four classes. The feature vectors extracted under four conditions in Table~\ref{tab:CV_Dataset_Classes} are used to train different classifiers, as demonstrated in Sec.~\ref{sec:ML_Classifier_Design_n_Analysis} for normality testing.

\subsection{ML Framework Design}
\label{ml_framework_design}

The ML framework to test the normality of I-V measurements is illustrated in Fig.~\ref{fig:IV_Profiles_ML_Framework}, which is an integral part of the electrochemistry workflow services (described in Sec.~\ref{subsec:Electrochemistry_Workflow}).
It consists of a pre-processing step for feature analysis, whose results are utilized by the ML model that validates the normality of I-V measurements. Based on its output, the domain computations are subsequently carried out using current I-V measurements (if found normal), or  an alert is generated for the examination by a human operator (otherwise). 
This ML framework expands the basic method in \cite{al-najjar-escience23} that used Ensemble of Trees (EOT) classifier trained with limited binary normal and abnormal data sets; specifically, it incorporates a fused-classifiers method that handles multiple classes (described in Section \ref{sec:ML_Classifier_Design_n_Analysis}).

We utilize this ML framework to study different classifiers and their accuracy to design an effective fused-classifiers model integrated into the electrochemistry workflow.

\section{ML Classifier Design and Analysis}
\label{sec:ML_Classifier_Design_n_Analysis}
\begin{figure*}[t]
\centering
\begin{subfigure}[c]{0.23\textwidth}
   \includegraphics[width=\textwidth,height=1.5in]{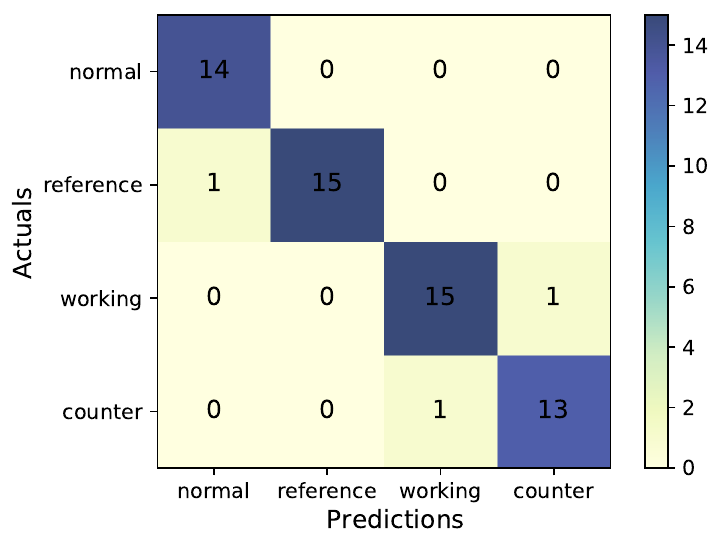}
\caption{CM: Extra Tree}
\label{subfig:CM_Extra_Tree}
\end{subfigure}
\hfill
\begin{subfigure}[c]{0.23\textwidth}
        \includegraphics[width=\textwidth,height=1.5in]{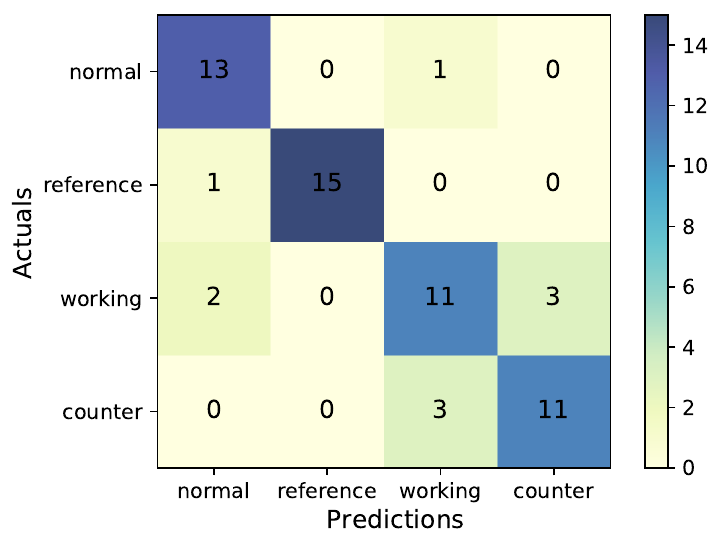}
\caption {{CM: EOT}}
\label{subfig:CM_EOT}
\end{subfigure}
\hfill
\begin{subfigure}[c]{0.23\textwidth}
   \includegraphics[width=\textwidth,height=1.5in]{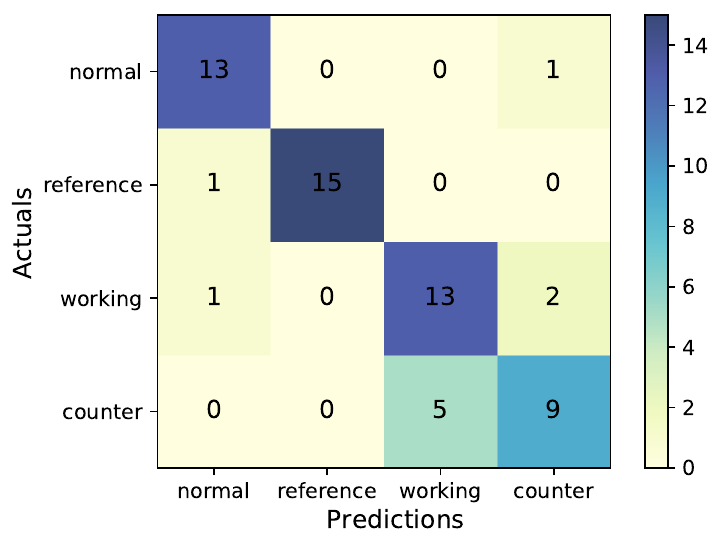}
\caption{{CM: Gaussian NB}}
\label{subfig:CM_Gaussian_NB}
\end{subfigure}
\begin{subfigure}[c]{0.23\textwidth}
        \includegraphics[width=\textwidth,height=1.5in]{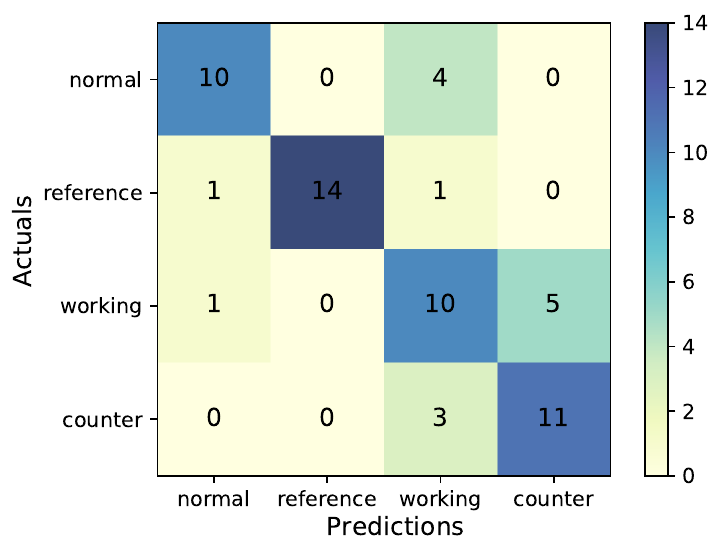}
\caption{{CM: Decision Tree}}
\label{subfig:CM_Decision_Tree}
\end{subfigure}
\hfill
\begin{subfigure}[c]{0.23\textwidth}
        \includegraphics[width=\textwidth,height=1.5in]{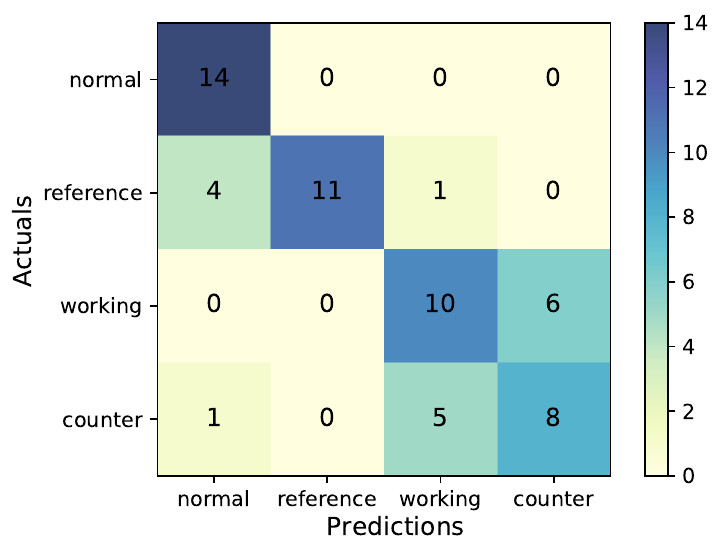}
\caption{{CM: Logistic Regression}}
\label{subfig:CM_Logistic_Regression}
\end{subfigure}
\hfill
\begin{subfigure}[c]{0.23\textwidth}
        \includegraphics[width=\textwidth,height=1.5in]{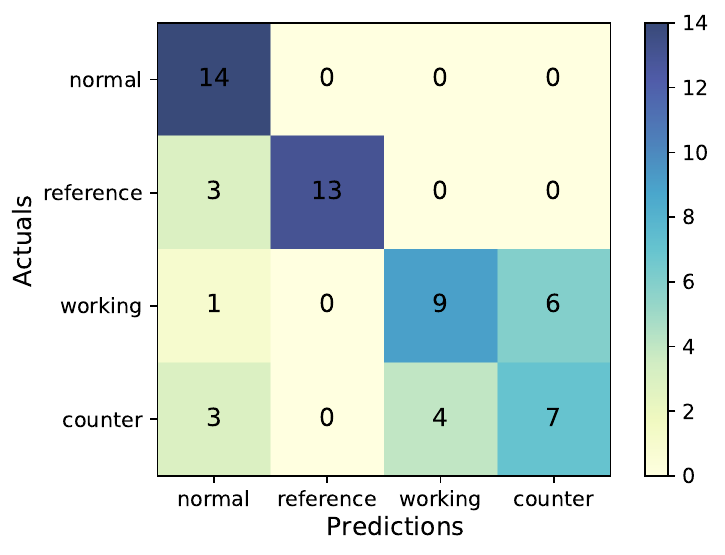}
\caption{{CM: SVM}}
\label{subfig:CM_SVM}
\end{subfigure}
\hfill
\begin{subfigure}[c]{0.23\textwidth}
        \includegraphics[width=\textwidth,height=1.5in]{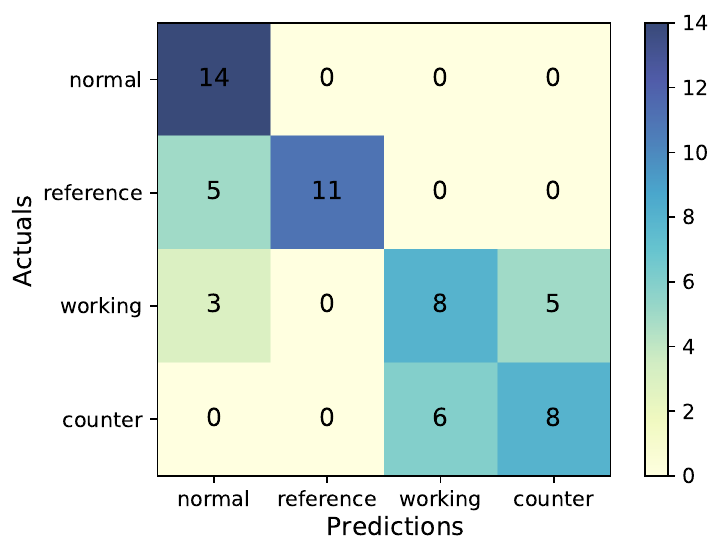}
\caption{{CM: KNN}}
\label{subfig:CM_KNN}
\end{subfigure}
\hfill
\caption{{Confusion matrices of multiple classifiers for CV datasets.}}
\label{fig:ml_CM_MC}
\vspace*{-0.2in}
\end{figure*}

\begin{figure*}[hbt]
\centering
\begin{subfigure}[c]{0.23\textwidth}
   \includegraphics[width=\textwidth,height=1.5in]{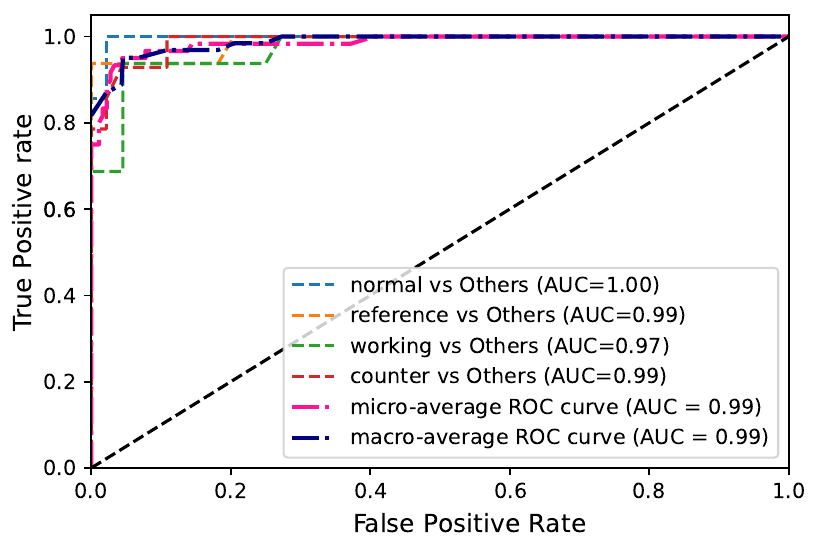}
\caption{ROC Curve: Extra Tree}
\label{subfig:ROC_Extra_Tree}
\end{subfigure}
\hfill
\begin{subfigure}[c]{0.23\textwidth}
        \includegraphics[width=\textwidth,height=1.5in]{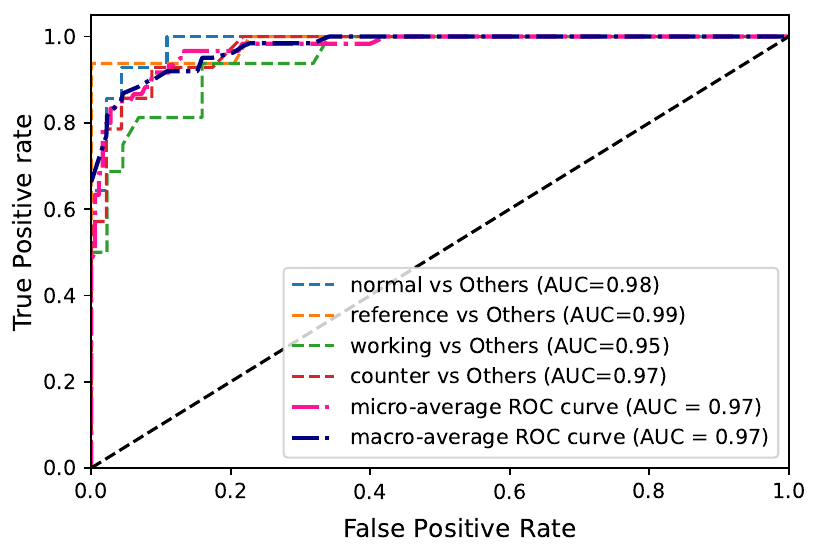}
\caption {{ROC Curve: EOT}}
\label{subfig:ROC_EOT}
\end{subfigure}
\hfill
\begin{subfigure}[c]{0.23\textwidth}
   \includegraphics[width=\textwidth,height=1.5in]{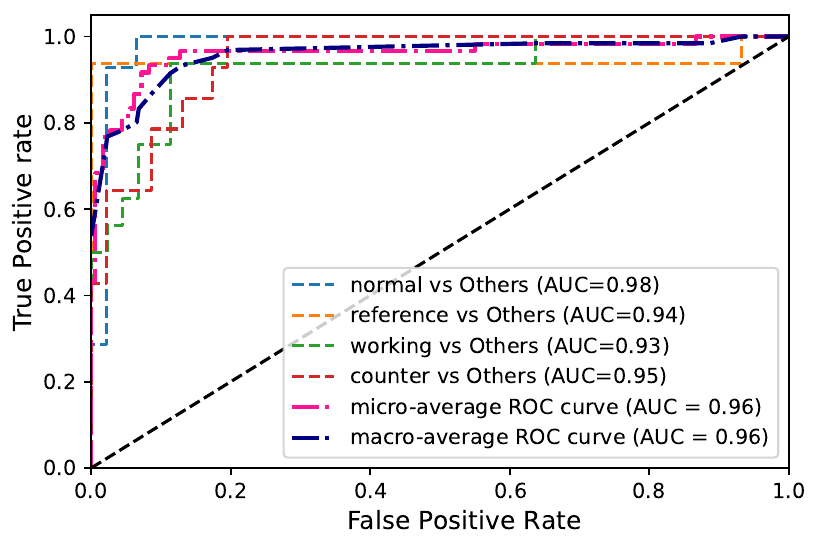}
\caption{{ROC Curve: Gaussian NB}}
\label{subfig:ROC_Gaussian_NB}
\end{subfigure}
\begin{subfigure}[c]{0.23\textwidth}
        \includegraphics[width=\textwidth,height=1.5in]{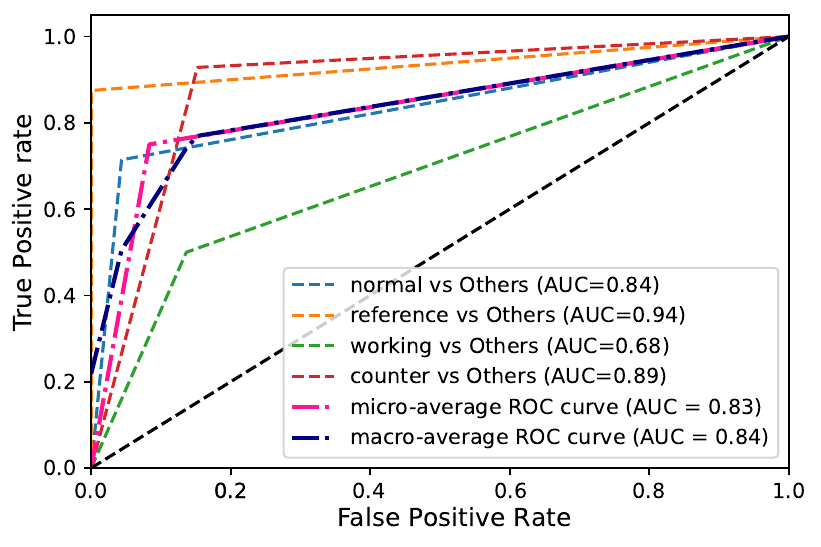}
\caption{{ROC Curve: Decision Tree}}
\label{subfig:ROC_Decision_Tree}
\end{subfigure}
\hfill
\begin{subfigure}[c]{0.26\textwidth}
        \includegraphics[width=\textwidth,height=1.5in]{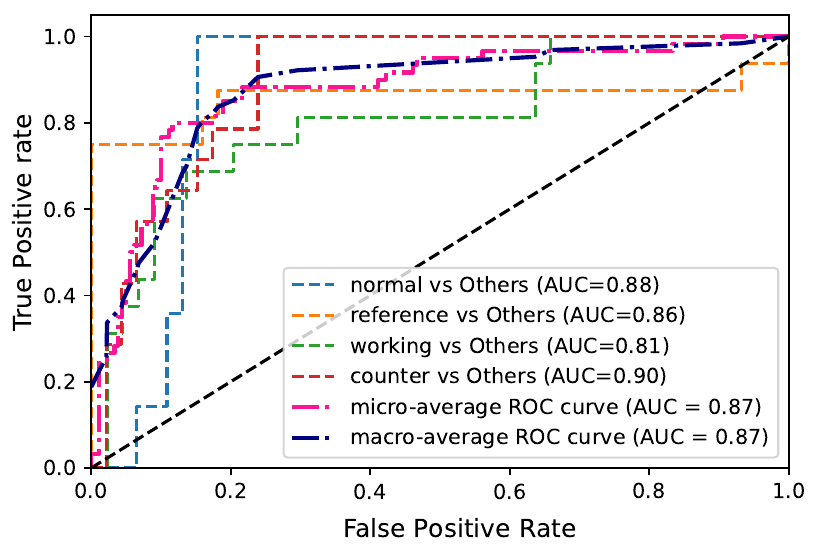}
\caption{{ROC Curve: Logistic Regression}}
\label{subfig:ROC_Logistic_Regression}
\end{subfigure}
\hfill
\begin{subfigure}[c]{0.23\textwidth}
        \includegraphics[width=\textwidth,height=1.5in]{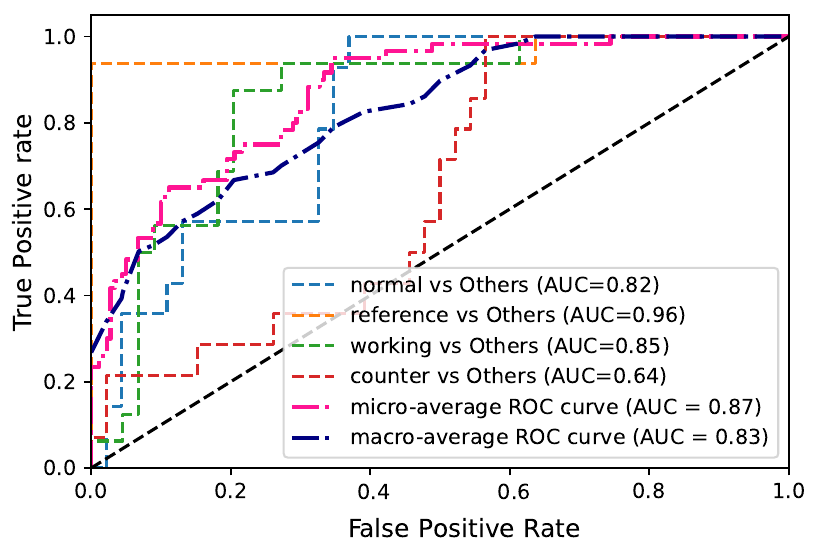}
\caption{{ROC Curve: SVM}}
\label{subfig:ROC_SVM}
\end{subfigure}
\hfill
\begin{subfigure}[c]{0.23\textwidth}
        \includegraphics[width=\textwidth,height=1.5in]{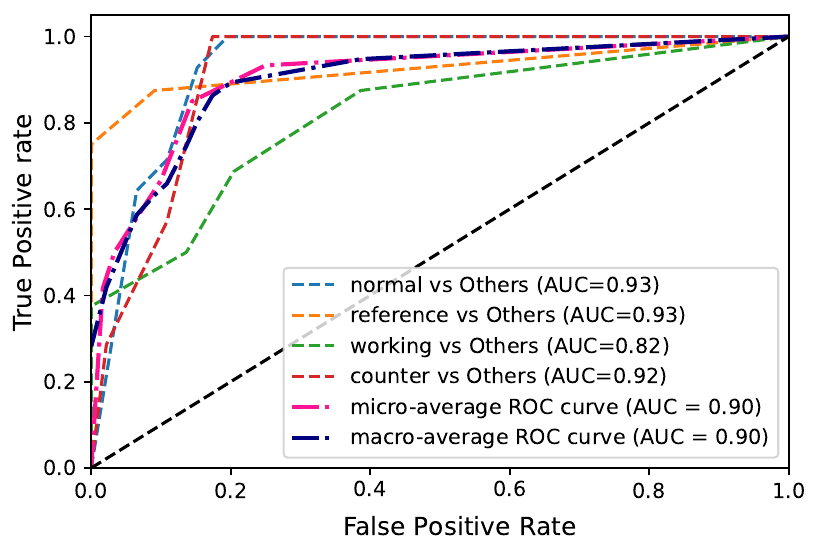}
\caption{{ROC Curve: KNN}}
\label{subfig:ROC_KNN}
\end{subfigure}
\hfill
\caption{{ROC curves of multiple classifiers for CV datasets.}}
\label{fig:ml_ROC_MC}
\end{figure*}

There is a wide variety of ML classification methods that can be applied to I-V features described in the previous section.
Indeed, the abundance of various ML methods makes their selection particularly hard since the classifier theory indicates that there is no single best method for finite samples \cite{DGL96}.
Hence, we carry out a systematic study of a variety of them as well as their fusers that combine the advantages of multiple classifiers, and design a fused-classifiers solution based on the performance on CV dataset and 
the generalization equations that analytically characterize the performance beyond training.
Scikit-learn Python package is used for investigating various ML classifiers using multiple performance metrics. 

\subsection{Multiple Classifiers}
\label{subsec:Multiple_classifiers}

We study multiple classifiers: Extra Trees (ET), EOT, Gaussian Naive Base (Gaussian NB), Decision Tree (DT), Logistic Regression (LR), Support Vector Machine (SVM),  and K-Nearest Neighbors (KNN).  They are chosen to represent the diversity of ML classifier designs, namely, 
 non-smooth (ET, EOT and DT), smooth (SVM),  statistical (NB and LR) and structural (KNN).
 
We assess their performance by estimating the confusion matrix, and the Receiver Operating Characteristic (ROC) curve which is a plot of the detection rate as a function of false alarm rate obtained by varying the classifier parameters. 
Estimates of these two quantities are shown in Figs.~ \ref{fig:ml_CM_MC} and
\ref{fig:ml_ROC_MC}, respectively, under 5-fold cross-validation of CV dataset, which show significant variations among the classifiers.
These metrics are succinctly summarized by the average accuracy and Area Under ROC Curve (AUC), which are estimated in   
Table~\ref{tab:Classifiers_performance}.
The ET and EOT have the highest accuracy and others are less than 85$\%$ accurate with the lowest 68$\%$ by KNN.

\begin{table}[htbp]
  \centering
  \caption{Classifiers performance to classify CV conditions}
    \begin{tabular}{|c|c|c|c|}
    \hline
    \rowcolor[rgb]{ .851,  .851,  .851} \textbf{No.}   & \textbf{Model} & \textbf{Accuracy(Avg.)} & \textbf{AUC} \\
    \hline
    \cellcolor[rgb]{ .851,  .851,  .851}1     & ET & 0.933 & 0.986 \\
    \hline
    \cellcolor[rgb]{ .851,  .851,  .851}2     & EOT & 0.85  & 0.973 \\
    \hline
    \cellcolor[rgb]{ .851,  .851,  .851}3     & Gaussian NB & 0.833 & 0.949 \\
    \hline
    \cellcolor[rgb]{ .851,  .851,  .851}4     & Decision Tree & 0.783 & 0.836 \\
    \hline
    \cellcolor[rgb]{ .851,  .851,  .851}5     & Logistic Regression & 0.717 & 0.86 \\
    \hline
    \cellcolor[rgb]{ .851,  .851,  .851}6     & SVM   & 0.717 & 0.817 \\
    \hline
    \cellcolor[rgb]{ .851,  .851,  .851}7     & KNN & 0.683 & 0.898 \\
    \hline
    \end{tabular}%
  \label{tab:Classifiers_performance}%
    \vspace*{-0.2in}
\end{table}%

\begin{figure}[t]
\begin{center}
\includegraphics[width=0.35\textwidth,height=2in]{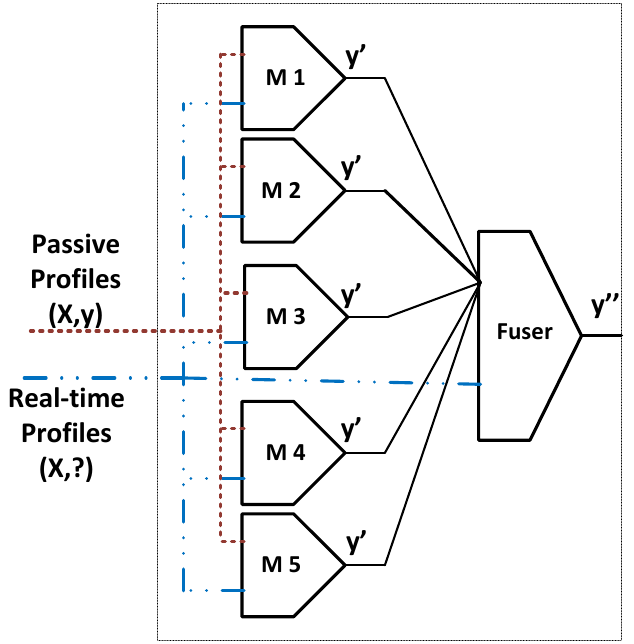}
\end{center}
\caption{Fuser design of multiple classifiers} 
\label{fig:Fuser_design}
\end{figure}

\subsection{Fusers of Multiple Classifiers}
\label{ML_classifiers}

\begin{figure}[t]
\centering
\begin{subfigure}[c]{0.245\textwidth}
\includegraphics[width=\textwidth,height=1.5in]{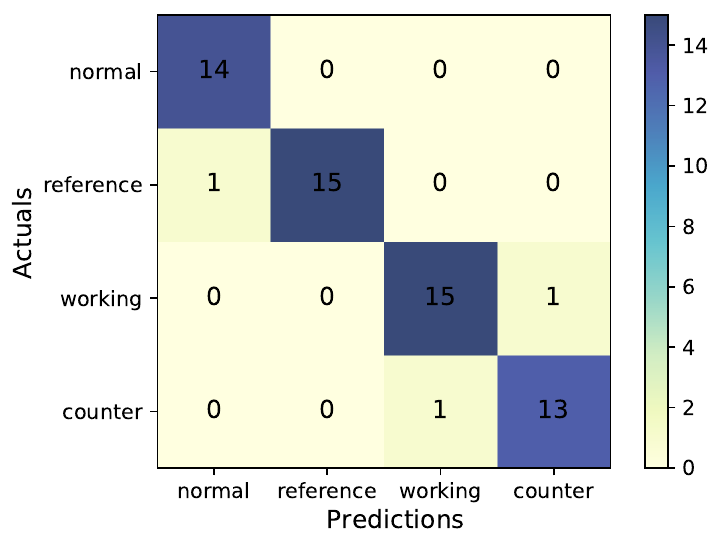}
\caption{{CM: EOT}}
\label{subfig:EOT_Confusion_Matrix}
\end{subfigure}
\hspace*{0.2em}
\begin{subfigure}[c]{0.225\textwidth}
\includegraphics[width=\textwidth,height=1.5in]{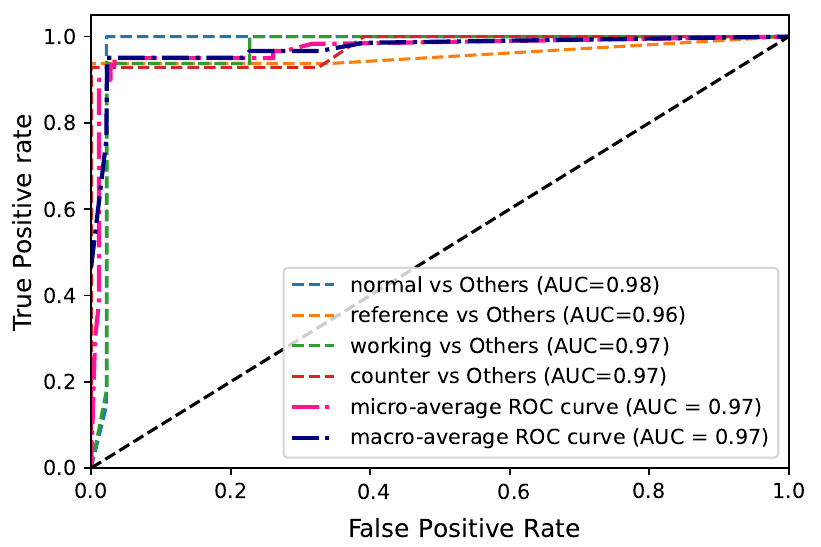}
\caption{{ROC Curve: EOT}}
\label{subfig:EOT_Roc_Curve}
\end{subfigure}
\caption{{EOT fuser performance analysis}}
\label{fig:EOT_fuzer_performance_analysis}
\vspace*{-0.2in}
\end{figure}

\begin{figure*}[t]
\centering
\begin{subfigure}[c]{0.23\textwidth}
        \includegraphics[width=\textwidth,height=1.5in]{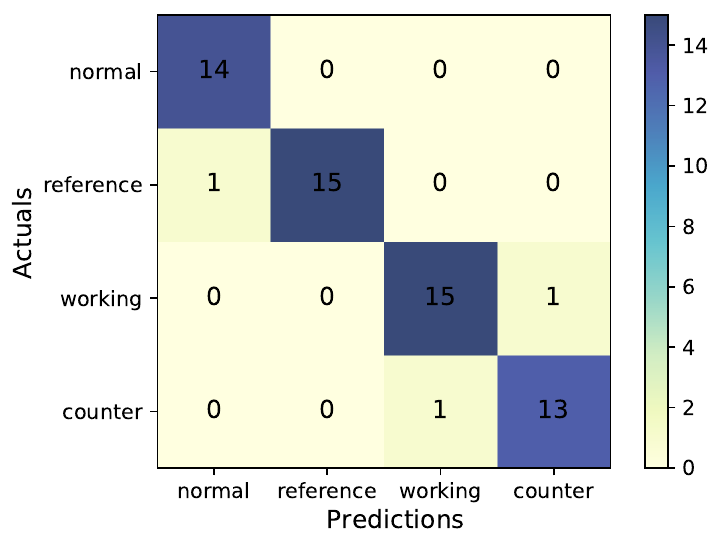}
\caption{{CM: fuser (Decision Tree)}}
\label{subfig:CM_F_Decision_Tree}
\end{subfigure}
\hfill
\begin{subfigure}[c]{0.23\textwidth}
   \includegraphics[width=\textwidth,height=1.5in]{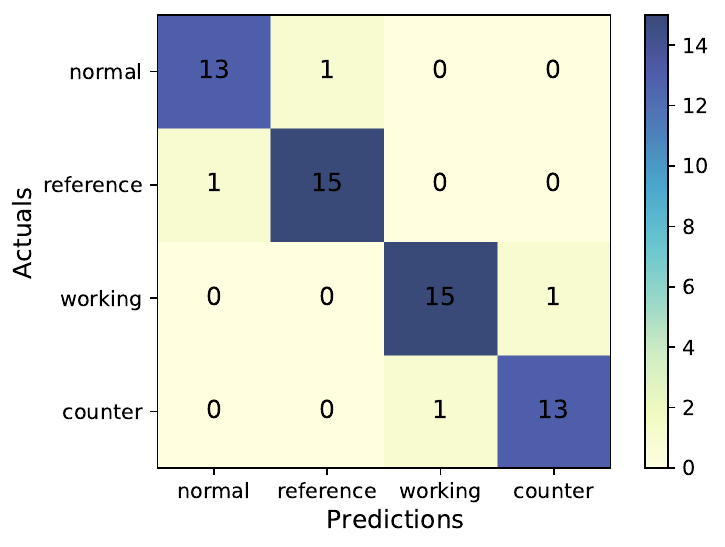}
\caption{{CM: fuser (Gaussian NB)}}
\label{subfig:CM_F_Gaussian_NB}
\end{subfigure}
\hfill
\begin{subfigure}[c]{0.26\textwidth}
        \includegraphics[width=\textwidth,height=1.5in]{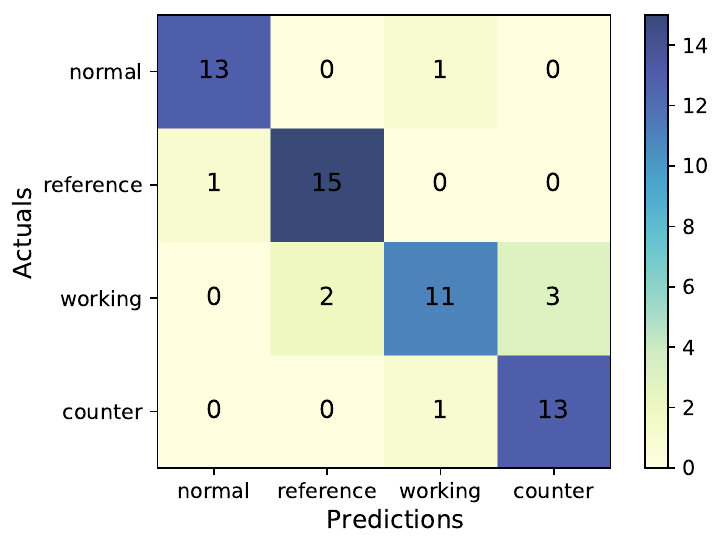}
\caption{{CM: fuser (Logistic Regression)}}
\label{subfig:CM_F_Logistic_Regression}
\end{subfigure}
\hfill
\begin{subfigure}[c]{0.23\textwidth}
   \includegraphics[width=\textwidth,height=1.5in]{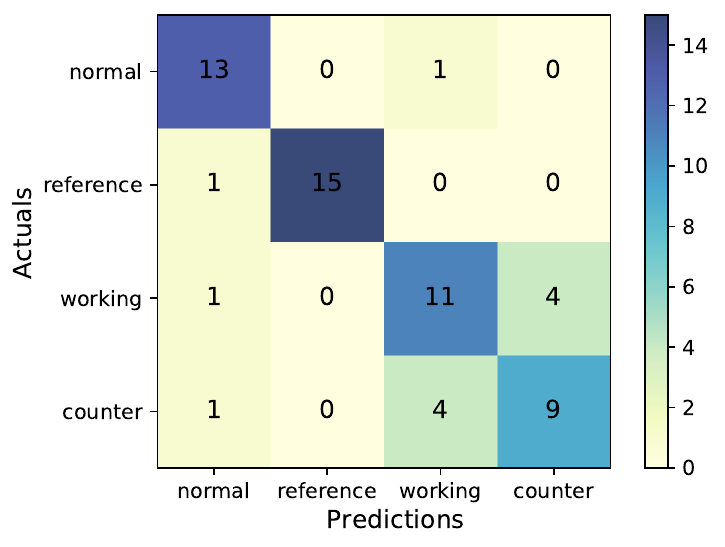}
\caption{CM: fuser (Extra Tree)}
\label{subfig:CM_F_Extra_Tree}
\end{subfigure}

\caption{{Confusion matrices of multiple fusers for CV dataset.}}
\label{fig:ml_CM_F}
\vspace*{-0.2in}
\end{figure*}

\begin{figure*}[t]
\centering
\begin{subfigure}[c]{0.245\textwidth}
        \includegraphics[width=\textwidth,height=1.5in]{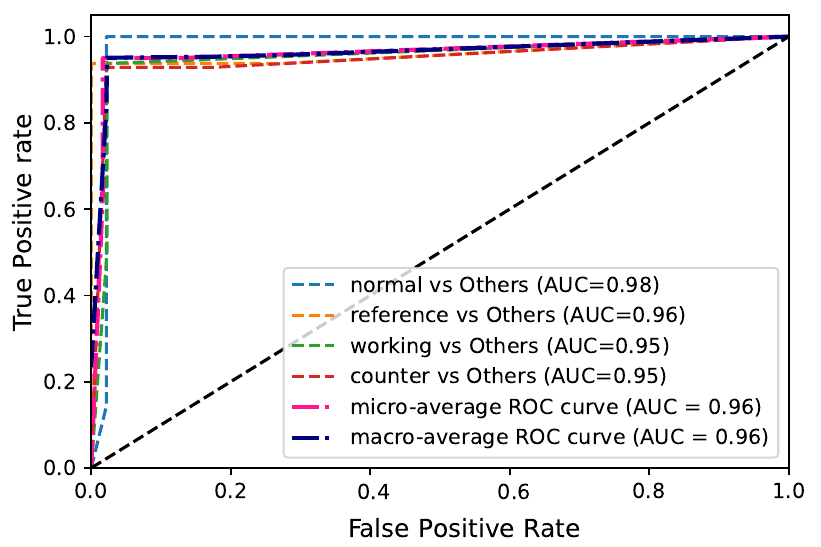}
            \centering \caption{{ROC Curve:\\ fuser (Decision Tree)}}
\label{subfig:ROC_F_Decision_Tree}
\end{subfigure}
\hfill
\begin{subfigure}[c]{0.245\textwidth}
   \includegraphics[width=\textwidth,height=1.5in]{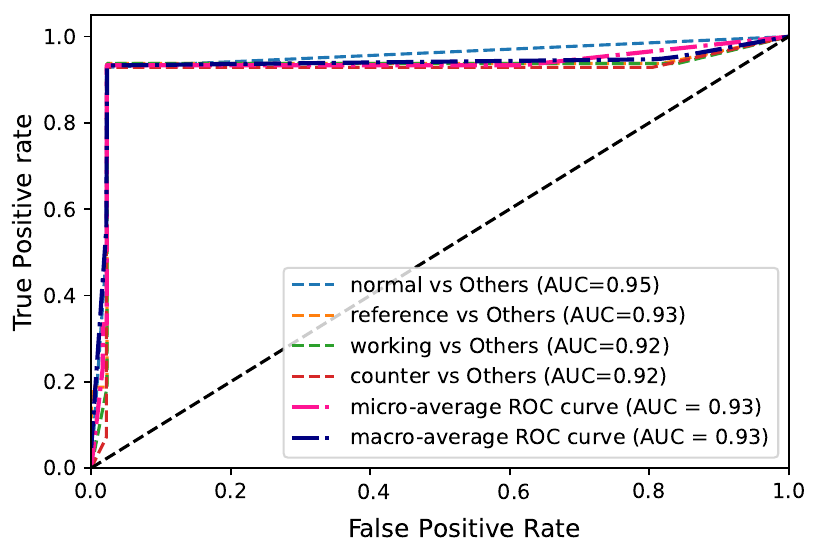}
        \centering \caption{{ROC Curve:\\ fuser (Gaussian NB)}}
\label{subfig:ROC_F_Gaussian_NB}
\end{subfigure}
\hfill
\begin{subfigure}[c]{0.245\textwidth}
        \includegraphics[width=\textwidth,height=1.5in]{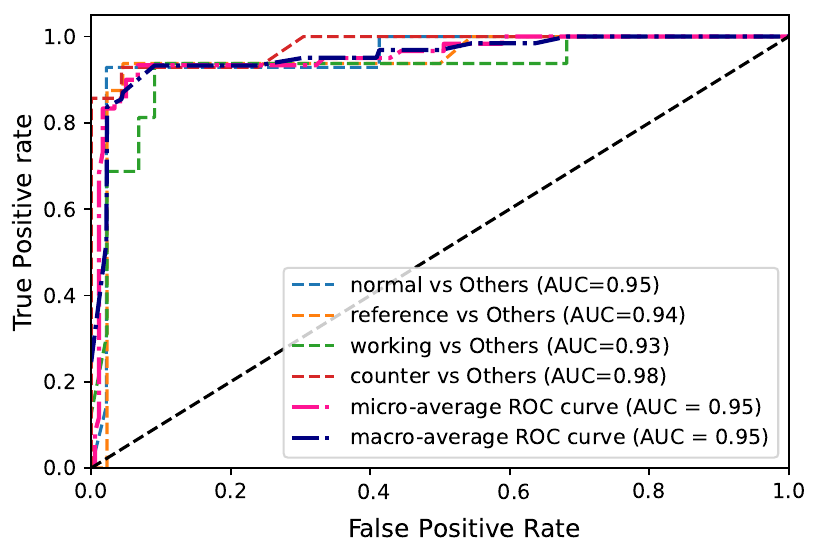}
            \centering \caption{{ROC Curve:\\ fuser (Logistic Regression)}}
\label{subfig:ROC_F_Logistic_Regression}
\end{subfigure}
\hfill
\begin{subfigure}[c]{0.245\textwidth}
   \includegraphics[width=\textwidth,height=1.5in]{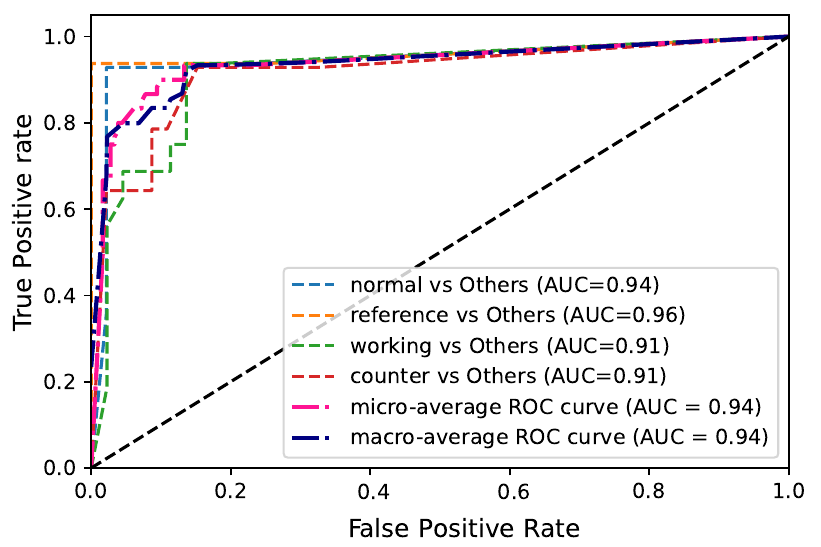}
        \centering \caption{ROC Curve:\\ fuser (Extra Tree)}       
\label{subfig:ROC_F_Extra_Tree}
\end{subfigure}

\caption{{ROC curves of multiple fusers based on CV dataset.}}
\label{fig:ml_ROC_F}
\vspace*{-0.2in}
\end{figure*}

We consider a fused-classifiers method, referred to simply as {\it fuser}, whose input is the outputs of its constituent classifiers and output is the classification, as shown in Fig.~\ref{fig:Fuser_design}. 
The fuser, which itself is a classifier, is trained with an input matrix of vectors of predicted classes of top five models and output of class labels.
The performance of top five fusers is shown in Table~\ref{tab:Fuser_performance}. 
Compared to single ML models, fusers generally have higher accuracy, with 95$\%$ for EOT fuser. The confusion matrix and ROC plots for EOT fuser in Fig.~\ref{fig:EOT_fuzer_performance_analysis} shows only three missed classes (Fig~\ref{subfig:EOT_Confusion_Matrix}), and has 0.97 AUC (Fig~\ref{subfig:EOT_Roc_Curve}).
The performance metrics of other four fused-classifiers with 5-fold cross-validation of the top five classifiers are shown in Table~\ref{tab:Fuser_performance}, which correspond to the confusion matrices (CM) and ROC curves shown in Figs.~\ref{fig:ml_CM_F} and \ref{fig:ml_ROC_F}, respectively. The performance metrics still show improvements of fusers compared with single classifiers shown in Figs.~\ref{fig:ml_CM_MC} and \ref{fig:ml_ROC_MC}.

In addition to performance metrics, the fusers incorporate advantages of different designs of their component classifiers, and  mitigate the over-fitting often found in single classifier designs.
Among the classifiers, ET has the highest accuracy but has lowest when used as a fuser.
On the other hand, EOT has lower accuracy as a classifier but has the highest accuracy as a fuser using others its classifiers.
Based on these results,  we integrated the EOT fuser with the top five classifiers in Table~\ref{tab:Classifiers_performance} into ML framework in Fig.~\ref{fig:IV_Profiles_ML_Framework}.

\begin{table}[htbp]
  \centering
  \caption{Fuser classification performance for CV dataset}
    \begin{tabular}{|c|c|c|c|}
    \hline
    \rowcolor[rgb]{ .851,  .851,  .851} \textbf{No.} & \textbf{Fuzer} & \textbf{Accuracy(Avg.)} & \textbf{AUC} \\
    \hline
    \cellcolor[rgb]{ .851,  .851,  .851} 1     & EOT & 0.95 & 0.97 \\
    \hline
    \cellcolor[rgb]{ .851,  .851,  .851} 2     & Decision Tree & 0.95 & 0.96 \\
    \hline
    \cellcolor[rgb]{ .851,  .851,  .851} 3     & Gaussian NB & 0.933 & 0.929 \\
    \hline
    \cellcolor[rgb]{ .851,  .851,  .851} 4     & Logistic Regression & 0.867 & 0.95 \\
    \hline
    \cellcolor[rgb]{ .851,  .851,  .851} 5     & ET & 0.8 & 0.933 \\
    \hline
    \end{tabular}%
  \label{tab:Fuser_performance}%
  \vspace*{-0.15in}
\end{table}%

\subsection{ML Generalization Equations}
\label{Generlization_Equations}

The proposed ML method, consisting of regression-based feature estimation and fused-classifiers detection, is analytically validated here. We derive the generalization equations for the underlying  regressions, classifiers and fusers, which have the same underlying analytical formulation \cite{Rao2021mfi,Rao2021fusion}.
In each case, the size of data set used for estimation is $l$;
for GPR regression it corresponds to the number of individual I-V measurements used for feature vector estimation, and for classifiers and fusers it corresponds to total number of normal and other I-V profiles used for training. 
Let $X$ and $Y$ correspond to input and output of an estimator $f$ whose generalization error is
$$ I (f) = \int L(Y,f(X)) d \mathbb{P}_{X,Y},$$ where $L(Y, f(X))$ is the error of estimate of $f(X)$ in predicting $Y$.
Here, $X$ and $Y$ are distributed according to complex, (mostly) unknown distribution $\mathbb{P}_{X,Y}$ that 
depends on data, measurement and other errors. 
For feature estimation, estimator $f_{GPR}$ correspond to regression function with $X$ and $Y$ corresponding to $V$ and $I$ measurements respectively.
For classifier $f_C$, $X$ corresponds to 10-d feature vector and $Y$ is  CV dataset class, and 
for fuser $f_F$, $X$ corresponds to the vector of outputs of its classifiers for a 10-d vector and $Y$ corresponds to CV dataset class.
The empirical error $\hat{I} (f)$ is a sample-based approximation of $I(f)$ which is minimized by estimator $\hat{f}$ within $\hat{\epsilon}$.
The generalization equation for $\hat{f}$ is
$$
\mathbb{P}^l_{X,Y} \left [ I ( \hat{f}) - I( f^*) > \epsilon \right ] < \delta \left ( \epsilon, \hat{\epsilon}, l \right ),
$$
where $\hat{f}$ is the GPR estimate or classifier/fuser, and $f^*$ is best possible estimator that minimizes the expected error $I(.)$.
This equation guarantees that the expected error of estimator $\hat{f}$ is within $\epsilon$ of optimal with confidence probability $1-\delta(.)$ that depends on $l$ and the training error $\hat{\epsilon}$ of $\hat{f}$. This performance guarantee holds irrespective of and without requiring the knowledge of $\mathbb{P}_{X,Y}$.

The generalization equations have been very useful in establishing the solvability and assessing the performance of ML methods \cite{Vapnik98,Murphy2020}, and in our case establish that both feature estimation and classification problems are solvable by respective ML methods.
For GPR, the confidence function is 
$ \delta_{GPR} = 8 \left ( \frac{32 \max(A,C)}{\epsilon} \right )^{2N_K} e^{- \epsilon^2l/512}, 
$
where $N_K$ is the number of component Gaussian functions and $A$ and $C$ are constraints that bound the estimators \cite{Raoetal2020ned}.
The confidence function of classifier or fuser (such as EOT and ET) is 
$ \delta_{C}  = 8 g \left (  1+ \frac{ 256B N_L}{\epsilon} \right ) e^{-\epsilon^2 l/2048} ,$ 
where $B$ is a bound and $N_L$ is the number of leaves of estimator \cite{Raoetal2020ned}; similar results are derived for other classifiers and fusers using the bounded total variation (details in \cite{Rao2021mfi}).
Thus, these generalization equations  analytically justify the proposed ML method, since their existence shows that the underlying problems are ML-solvable \cite{Rao2021fusion}.


\section{Conclusion and Future work}
\label{sec:conclusions}
We presented an electrochemistry computing 
 platform with networking and software eSolutions for incorporating a chemical synthesis workstation and a delivery mobile robot into ICE to support automated  workflows. 
We described a workflow that generates I-V measurements of an electrochemical solution in the reactor cell connected to a potentiostat via electrodes.
We developed and deployed an ML framework as a part of this workflow for ensuring that I-V measurements are consistent with the data expected for a standard electrochemical cell operating under normal conditions, and detect abnormal conditions, such as disconnected electrodes.
The ML framework employs GPR-based feature estimation method, and a fused-classifiers method that combines the benefits of a variety of classifier methods. 
This method is justified by studying a variety of ML classifiers using experimental datasets, and deriving the generalization equations of feature estimation and classification methods that ensure performance beyond training.

There are several future research areas that can be pursued.
This work is primarily focused on ICE infrastructure development, and it would be of interest to study the performance and productivity in term of electrochemistry and scientific productivity gains.
The expansion of ML method to include additional failures, including ones based on measurements collected during the production and delivery of compounds, will be of future interest.
It would also be of interest to derive more customized versions of the generalization equations by incorporating the detailed parameters derived from electrochemistry considerations. Exploring the resilience and fault tolerance of cross-facility electrochemistry workflows and autonomous orchestration of electrochemistry workstations dispersed across different science facilities, are of future interest.

\bibliographystyle{ieeetr}
\bibliography{./ref.bib}

\end{document}